\documentclass[a4paper,11pt]{article}
\pdfoutput=1 
\usepackage[usenames,dvipsnames]{xcolor}
\usepackage{jheppub} 
\usepackage{enumerate}
\usepackage[T1]{fontenc} 
\usepackage{comment,epigraph}
\usepackage{physics}
\usepackage{float,soul,color}
\usepackage{dsfont,comment}
\usepackage{caption}
\usepackage{slashed}
\usepackage[mathscr]{euscript}
\usepackage{xhfill}
\usepackage{smartdiagram}
\usepackage{subcaption}
\usepackage{mathtools}
\usetikzlibrary{shapes.geometric}
\usetikzlibrary{arrows}
\usesmartdiagramlibrary{additions} 
\usepackage{gensymb}
\usepackage{booktabs}
\usepackage[normalem]{ulem}
\usepackage[bottom]{footmisc}
\usepackage[dvipsnames]{xcolor}
\usepackage{cancel}
\usepackage{bbm}
\usepackage{graphicx}
\usepackage{subcaption}

\definecolor{cobalt}{rgb}{0.0, 0.28, 0.67}

\newcommand{\ch}[1]{\textcolor{Purple}{}}

\title{\Huge \centering \textbf{Holographic supersolids}}

\author[a,b]{Matteo Baggioli}
\author[a]{, Giorgio Frangi}

\affiliation[a]{Wilczek Quantum Center, School of Physics and Astronomy, Shanghai Jiao Tong University, 800 Dongchuan Road, Minhang District, Shanghai 200240, China.}
\affiliation[b]{Shanghai Research Center for Quantum Sciences, Shanghai 201315.}

\emailAdd{b.matteo@sjtu.edu.cn}
\emailAdd{fgiorgio@sjtu.edu.cn}

\abstract{A supersolid is a system that presents long-range order and shear rigidity as a solid but which also supports a non-dissipative superflow as a superfluid. From an effective perspective, supersolids are identified with phases of matter that break spontaneously translational invariance together with a global $U(1)$ symmetry. By using this symmetry prescription, we build a holographic bottom-up model for supersolids and we start the investigation of its thermodynamic and mechanical properties. More precisely, we analyze the behaviour of the critical temperature, the condensate, the shear modulus and the viscosity across all the phase diagram. Finally, we successfully compare our results with a simple Ginzburg-Landau model for supersolids deriving some universal physical correlations between the observables mentioned above.}

\begin{document} 
\maketitle
\flushbottom

\section{Introduction}
 Crystalline solids\footnote{We will ignore in this manuscript the possibility of having solids lacking long-range order -- amorphous solids (e.g. glasses).} are characterized by the presence of long-range order and they exhibit macroscopic shear rigidity \cite{chaikin_lubensky_1995}. From a more fundamental perspective, solids correspond to ground states which spontaneously break translational invariance (usually down to a discrete group defined by the lattice vectors) and whose low-energy dynamics is controlled by the corresponding Goldstone modes -- the phonons \cite{Leutwyler:1996er}. In a similar fashion, superfluids are identified with ground states which support a non-dissipative flow with zero viscosity -- the superflow \cite{RevModPhys.71.S318}. Superfluid states break spontaneously a $U(1)$ global symmetry \cite{Son:2002zn,Nicolis:2011cs} and their low-energy physics is dominated by a Goldstone mode as well, accounting for the propagation of second sound \cite{doi:10.1063/1.3248499}.\footnote{More precisely, second sound arises as a combination of the $U(1)$ Goldstone mode and the original charge diffusion mode. The same pattern holds for a solid in which the phonons are not exactly the Goldstone excitations but a combination of them with momentum and energy fluctuations.}
 
A supersolid is a phase of matter that displays simultaneously the distinctive features of solids and superfluids \cite{RevModPhys.84.759,Balibar2010}. How can a rigid body with "infinite" viscosity flow freely as a superfluid? The questions of whether such a system could exist, how the superflow could appear in a solid and where a supersolid could be experimentally observed have occupied the minds of several physicists in the past seventy years \cite{PhysRev.106.161,GROSS195857,RevModPhys.34.694,PhysRevA.2.256,PhysRevLett.25.1543} since the original observation by O. Penrose and Onsager \cite{PhysRev.104.576}. Despite the tremendous theoretical progress, supersolid systems have eluded clear detection for the longest time, and it was in recent times only that they have finally been observed in laboratories: historically, supersolid behavior has been first looked for in solid helium \cite{doi:10.1126/science.1196409,kim2004probable,kim2004observation,rittner2006observation,PhysRevLett.109.155301,kim2014upper}, while in the last few years advancements in the field of ultracold atom gases provided some clear signatures of it \cite{doi:10.1126/science.aba4309,Guo2021,Norcia2021}. Nonetheless, a complete understanding of this phenomenon is still lacking.
 
From an abstract point of view, a supersolid is a state which spontaneously breaks translational (and rotational) invariance together with a $U(1)$ global symmetry. This will be our operative, and indeed agnostic in a microscopic sense, definition of a supersolid. In this work, we will limit ourselves to analyzing the possible low energy dynamics and physics of such a hypothetical system \cite{PhysRevLett.104.075302,doi:10.1063/1.2883895,Syshchenko_2009,Day2007,prokof2007makes,lin2009heat,anderson2008bose}. In the past decades, a lot of effort has been devoted to this direction, especially into finding the correct low-energy effective description and hydrodynamic framework for supersolids \cite{PhysRevB.53.5670,Saslow2012,hofmann2021hydrodynamics,PhysRevLett.97.125302,PhysRevLett.96.055301,PhysRevLett.94.175301,RevModPhys.46.705}. Recently, supersolids have been also considered as possible ground states in low-energy effective field theories with spontaneously broken Lorentz invariance \cite{Nicolis:2013lma,Delacretaz:2014jka,LVEFT3,Krichevsky:2020ury,Celoria:2017bbh}, which will be closer in spirit to our approach.
 
 Inspired by these ideas and questions, we will reformulate the problem of supersolids and their low-energy description in terms of holographic methods \cite{zaanen2015holographic,Natsuume:2014sfa,Hartnoll:2016apf,Baggioli:2019rrs}. Superfluid states have been successfully incorporated in the holographic framework more than ten years ago, using a simple abelian Higgs model \cite{Hartnoll:2008vx,Hartnoll_2008} (for a review of this model see \cite{Cai:2015cya}) and the corresponding low energy spectrum has been successfully matched to the predictions of relativistic superfluid hydrodynamics \cite{Herzog:2011ec,Schmitt:2014eka} in \cite{Arean:2021tks}.\footnote{The hydrodynamics description and the agreement with the holographic model has been also recently extended to the case of softly broken $U(1)$ symmetry \cite{Ammon:2021slb,Donos:2021pkk}.} Holographic solids, i.e. holographic systems spontaneously breaking translations, have attracted a lot of interest recently as possible toy-models to understand the complex dynamics of strongly correlated materials such as strange metals. A very convenient scenario is obtained by imposing large scale homogeneity and avoiding any spatial dependence in the stress tensor and related quantities. Models of this sort are usually referred to as holographic homogeneous solids \cite{baggioli2015electron,Baggioli:2021xuv,Donos:2013eha,Grozdanov:2018ewh,Nakamura:2009tf,Amoretti:2017frz} and have represented a very efficient platform to perform computations in a controllable way. The simplest among the homogeneous setups is the holographic axion model \cite{Baggioli:2021xuv}, which presents a well-defined elastic response together with the presence of propagating phonon modes \cite{Alberte:2016xja,Alberte:2017oqx,Baggioli:2019abx}. As in the case of superfluids, its low energy dynamics is well described by hydrodynamics with broken translations \cite{Armas:2019sbe,Ammon:2020xyv,Armas:2020bmo,Baggioli:2020edn}. 
 
 In this work, we combine the holographic superfluid model of \cite{Hartnoll:2008vx} with the holographic solid model of \cite{baggioli2015electron} to build a holographic supersolid -- a holographic system which breaks spontaneously translations together with a global $U(1)$ symmetry\footnote{See \cite{Kiritsis:2015oxa} for a previous attempt of describing supersolids using holographic superfluid models with momentum relaxation.\color{black}}. The combination of superfluidity with broken translations has been already discussed in the holographic literature \cite{Baggioli:2015zoa,Baggioli:2015dwa,Zeng:2014uoa,Ling:2014laa,Andrade:2014xca,Erdmenger:2015qqa,Kim:2015dna,Jeong:2021wiu,Ling:2016lis,Kim:2016jjk,KimHomes,Kiritsis:2015hoa} but always in the context of momentum dissipation, i.e. explicit breaking of translations. A crucial novel ingredient is therefore to allow for translations to be broken spontaneously and give rise to the rigidity of our holographic system. \medskip
 
 The manuscript is organized as follows. In Section \ref{effectivesection} we describe a simple Ginzburg-Landau (GL) theory for supersolids which will serve as our theoretical background; in Section \ref{section:model} we introduce the holographic model for supersolids used in this work; in Section \ref{sec:results} we present our main results and characterize in detail the thermodynamic and mechanical properties of the holographic model, comparing it with the GL theory mentioned above; finally, in Section \ref{sec:conclusions} we conclude and provide some thoughts for the future. Appendices \ref{app1}-\ref{app2} are left for further details about the theory of nonlinear elasticity and for technicalities regarding the holographic model; Appendix \ref{app3} provides some additional results that are not discussed in the main text.

\subsubsection*{Note added}
This work is a direct continuation of the Master thesis of Giorgio Frangi "\textit{A holographic study of a two-dimensional superconductor under strain deformations}" [2021, Università degli studi di Milano].

\section{Ginzburg-Landau theory of supersolids}
\label{effectivesection}
In this Section, we review the simple effective description of supersolids presented in \cite{PhysRevLett.96.055301} and its main features. For a more complete description of the low-energy physics and hydrodynamics of supersolids we refer to \cite{PhysRevB.53.5670,Saslow2012,hofmann2021hydrodynamics,PhysRevLett.97.125302,PhysRevLett.96.055301,PhysRevLett.94.175301,RevModPhys.46.705}.

\begin{figure}
    \centering
    \includegraphics[trim=25 0 0 0,clip,width=0.7 \linewidth]{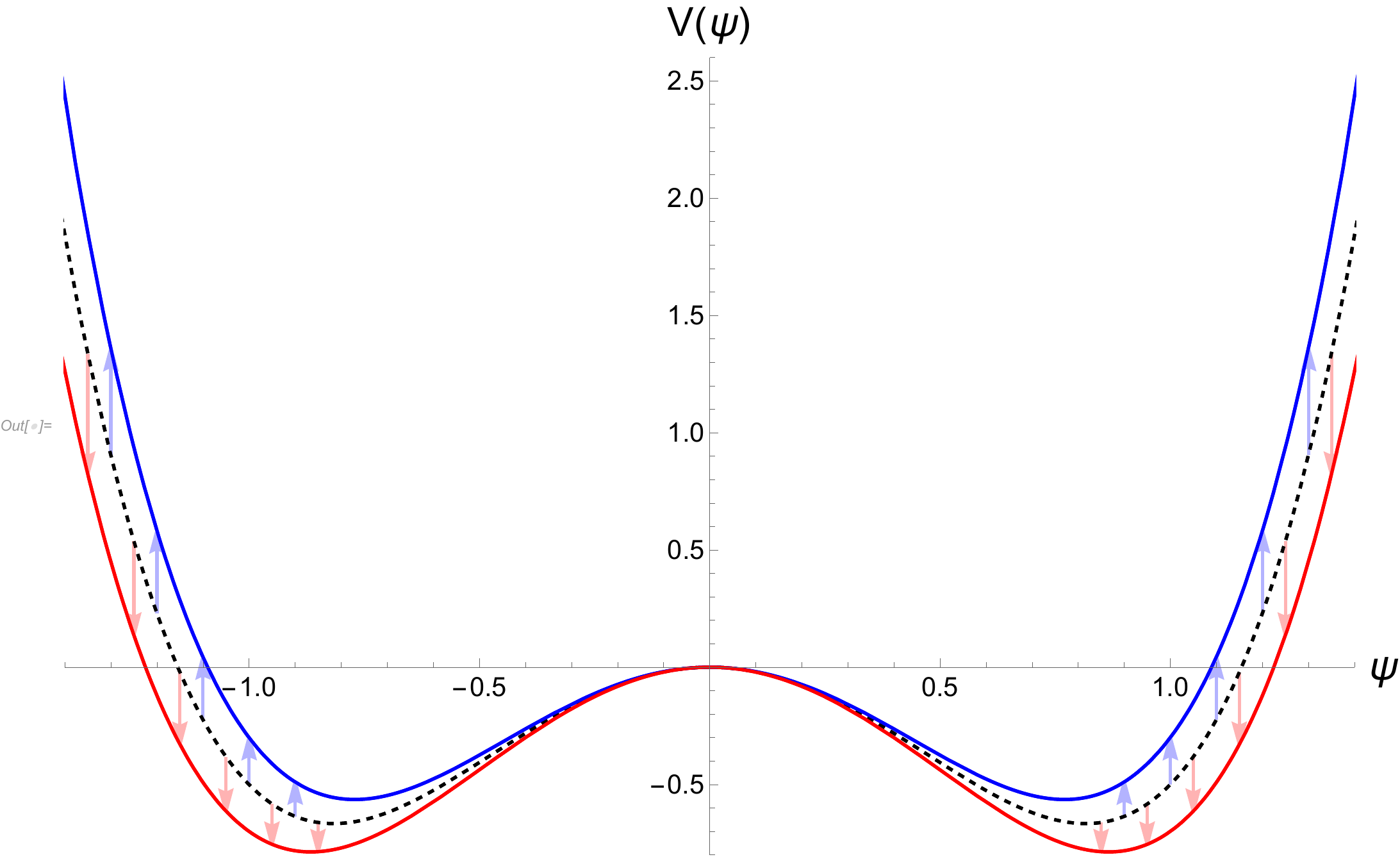}
    \caption{Effect of small, homogeneous deformations $(\delta = \pm 0.1, \, \epsilon = 0)$ on a slice of the Mexican hat potential appearing in \eqref{eqsupersolid}, at a given temperature $T$. We start with assuming $a_s(T) = -2$ and $b_s(T) = 1.5$ (black, dashed curve), then apply \eqref{fullTc} and its analogous expression for $b_s(T)$, neglecting $\delta^2$ terms and taking $a_1(T) = b_1(T) = 1$. For this choice of coefficients, a compression (red, $\delta < 0$) enhances the condensate, whilst a dilatation (blue, $\delta > 0$) weakens it.}
    \label{fig:add2}
\end{figure}

The phenomenology of an isotropic $s$-wave superfluid near its transition can be captured via the Ginzburg-Landau (GL) formalism by writing the following free energy functional \cite{Schmitt:2014eka}:
\begin{equation}\label{GLeq}
    F_s[\Psi(x)]=\int d^d x \left[\frac{1}{2}\,|\partial \Psi|^2+ a_s(T) |\Psi|^2+ b_s(T) |\Psi|^4\,+\,\dots\right],
\end{equation}
where $\Psi$ is the (complex) scalar order parameter and $d$ is the number of spatial dimensions of a flat space. This expression, usually referred to as the Mexican hat potential (see Fig.\ref{fig:add2}), has to be taken as a low-energy effective description valid close to the phase transition, where $\langle \Psi \rangle \ll 1$. The ellipsis refers to possible higher order corrections, both in $\Psi$ and its derivatives. The functional in Eq.\eqref{GLeq} is clearly invariant under a global $U(1)$ transformation $\Psi \rightarrow e^{i \varphi} \, \Psi$.

Whenever $\langle \Psi \rangle \neq 0$, the ground state of the system breaks spontaneously the original $U(1)$ global symmetry and the system is in a superfluid phase. This state is often called the broken phase, as opposed to the normal phase, $\langle \Psi\rangle = 0$, in which the $U(1)$ symmetry is intact. The transition from one phase to the other is introduced phenomenologically by imposing that the coefficient $a_s(T)$ changes sign at some critical temperature $T_c^{(0)}$, becoming negative below it and shifting the minimum of the free energy away from the unbroken ground state $\langle \Psi \rangle =0$. Here, $T_c^{(0)}$ is used to indicate the value of the critical temperature in the purely superfluid state, where translations are not broken in any way. The additional requirement $b_s(T) > 0$ ensures that the argument of the functional in Eq.\eqref{GLeq} is bounded from below and consequently the system is stable.

The $a_s(T)$ parameter in the GL functional \eqref{GLeq} vanishes linearly at the critical temperature:
\begin{equation}
    a_s(T)= a_0\, \left(T-T_c^{(0)}\right)\,+\dots\,,
\end{equation}
below which, if the condition $a_0 >0$ is fulfilled, it acquires a negative value, causing the condensation. \medskip

In a similar way, it is possible to write down the free energy of an elastic medium deformed away from its equilibrium configuration as \cite{chaikin_lubensky_1995}:
\begin{equation}
    F_e[u] = \int d^d x\, \frac{1}{2}\,C_{\alpha \beta \gamma \delta} \,u_{\alpha \beta}\,u_{\delta \gamma}\,+\,\dots\,, \label{solid}
\end{equation}
with $u$ being the linear strain tensor, i.e. the symmetrized derivative of the displacement vector $u_\alpha = x'_\alpha - x_\alpha$, and $C$ the elastic tensor. Here, $x_\alpha$ is the position of the atoms in the initial configuration and $x'_\alpha$ the one after the mechanical deformation. Indices refer only to the spatial directions. For isotropic materials, $C$ can be decomposed as:
\begin{equation}\label{isotropic_eltensor}
    C_{\alpha\beta\gamma\delta} = \mathcal{K}\,\delta_{\alpha\beta}\delta_{\gamma\delta} + \mathcal{G} \left(\delta_{\alpha\gamma}\delta_{\beta\delta} + \delta_{\alpha\delta}\delta_{\beta\gamma}
    - \frac{2}{d}\, \delta_{\alpha\beta}\delta_{\gamma\delta} \right),
\end{equation}
with $d$ the number of spatial dimensions. $\mathcal{G}$ and $\mathcal{K}$ are respectively called \textit{shear} and \textit{bulk modulus}, since they are related to volume-preserving and shape-preserving deformations. \medskip

Given this premise, the simplest way to think of a supersolid in its low-energy regime is to combine the superfluid free energy \eqref{GLeq} with the solid one \eqref{solid}, by making the phenomenological coefficients of the former analytic functions of the linear strain tensor $u_{\alpha\beta}$. In this picture, the low-energy degrees of freedom are given by the phonons, together with the superfluid Goldstone mode. At quadratic order in $u_{\alpha\beta}$, the coupling between the two types of degrees of freedom can be reconstructed via the following prescription:
\begin{equation}\label{eq:expansion}
    \begin{cases}
      a_s(T) \longrightarrow a(T,u) = a_s(T) + {a_1(T)}_{\alpha\beta} u_{\alpha\beta} + \frac{1}{2}{a_2(T)}_{\alpha\beta\gamma\delta}u_{\alpha\beta}u_{\gamma\delta} + O(u^3) \\
      b_s(T) \longrightarrow b(T,u) = b_s(T) + {b_1(T)}_{\alpha\beta} u_{\alpha\beta} + \frac{1}{2}{b_2(T)}_{\alpha\beta\gamma\delta}u_{\alpha\beta}u_{\gamma\delta} + O(u^3)
    \end{cases} \, .
\end{equation}
Notice that, because the background metric is flat, at this point the position of the indices is not important. All in all, the total free energy of a supersolid (ss) is given by:
\begin{equation}
\begin{split}
    F_{ss}[\Psi,u]= F_s[\Psi] + F_e[u] + \int d^d x \, & \left( a_1(T) |\Psi|^2 + b_1(T) |\Psi|^4 \right)_{\alpha\beta} u_{\alpha\beta} + \\ & + \frac{1}{2} \left( a_2(T) |\Psi|^2 + b_2(T) |\Psi|^4 \right)_{\alpha\beta\gamma\delta} u_{\alpha\beta} u_{\gamma\delta}\,.
\end{split}
\label{eqsupersolid}
\end{equation}
Before analyzing in detail the physics hidden in Eq. \eqref{eqsupersolid}, let us notice that, due to isotropy, the tensorial coefficients of the expansion \eqref{eq:expansion} can be decomposed in a way analogous to the one reported in \eqref{isotropic_eltensor}:
\begin{gather}\label{eq:isotropy}
    a_1(T)_{\alpha\beta} = a_1(T)\, \delta_{\alpha\beta}\,, \\
    a_2(T)_{\alpha\beta\gamma\delta} = a_2^{(\mathcal{K})}(T) \delta_{\alpha\beta}\delta_{\gamma\delta} + a_2^{(\mathcal{G})}(T) \left(\delta_{\alpha\beta}\delta_{\gamma\delta} + \delta_{\alpha\gamma}\delta_{\beta\delta} - \frac{2}{d} \delta_{\alpha\delta}\delta_{\beta\gamma} \right).
\end{gather}
A similar decomposition holds for $b$ coefficients as well.

In the remainder of this work, we restrict our attention to \textit{homogeneous} deformations of isotropic media with two spatial dimensions ($d = 2$). In Appendix \ref{app1}, we show with rigorous symmetry arguments that homogeneity allows us to parameterize our deformations of interest with the following strain tensor:
\begin{equation}\label{eq:parameterization}
    u_{lin} = \frac{1}{2} \begin{pmatrix}
    \delta & \epsilon \\
    \epsilon & \delta
    \end{pmatrix}.
\end{equation}
The subscript emphasizes that the strain tensor in the equation above is the linear one which is valid only for small deformations, i.e $\delta,\,\epsilon \ll 1$.\footnote{In the following, we will distinguish it from the non-linear Lagrangian strain tensor which will be indicated with $E_{ij}$.} This justifies the expansions in $\delta$, $\epsilon$ performed in the next subsections.\\

If $\epsilon=0$, $\delta$ describes a purely volumetric deformation with  $\Delta V/V = \delta$\footnote{This definition will be kept also at the nonlinear level in the following.}, while in the opposite case ($\delta=0$), $\epsilon$ gives rise to a purely deviatoric term which modifies the shape and the angles in the medium without changing its volume.\footnote{At linear level $\epsilon/2$ is both the off-diagonal component of the strain tensor and of the deformation gradient. At nonlinear level, the two definitions will no longer coincide. We will keep defining $\epsilon$ as the off-diagonal component of the deformation gradient.} Homogeneity implies that the superfluid condensate $\Psi$ is constant in space as well.\medskip

We now review the main effects that the coupling of elasticity and superfluidity produces in this framework.

\subsubsection*{Critical temperature}
The first important consequence of coupling the solid and superfluid sectors into the supersolid effective action \eqref{eqsupersolid} is that the value of the original critical temperature $T_c^{(0)}$ is modified by the presence of a mechanical deformation.
In particular, using \eqref{eq:isotropy} and \eqref{eq:parameterization}, one finds that the corrected coefficient of the quadratic term $|\Psi|^2$ is now given by:
\begin{equation}
    a(T,u)=a_s(T) + a_1(T)\,\delta + a_2^{(\mathcal{K})}(T) \, \delta^2 +  a_2^{(\mathcal{G})}(T) \, \epsilon^2\,, \label{fullTc}
\end{equation}
so that it depends explicitly on the mechanical deformation parameters $\delta,\,\epsilon$.
It is then straightforward to obtain the corrections to the superfluid critical temperature $T_c^{(0)}$ induced by the presence of a solid component in the medium. By identifying the new critical temperature $T_c$ as the zero of the corrected quadratic coefficient $a(T,u)$ in Eq. \eqref{fullTc}, we find (respectively, for a purely deviatoric and a purely volumetric deformation):
\begin{align}
    \label{eq:shear_tc_corr}
    &\delta T_c^{(\epsilon)} = T_c - T_c^{(0)} = - \frac{a_2^{(\mathcal{G})}\left(T_c^{(0)}\right)}{a_0\left(T_c^{(0)}\right)} \, \epsilon^2\,,\\
    & \delta T_c^{(\delta)} = T_c - T_c^{(0)} =  - \frac{a_1\left(T_c^{(0)}\right)}{a_0\left(T_c^{(0)}\right)} \, \delta - \frac{a_2^{(\mathcal{K})}\left(T_c^{(0)}\right)}{a_0\left(T_c^{(0)}\right)} \, \delta^2\,.\label{eq:delta_tc_corr}
\end{align}
Importantly, we already observe that the correction to the critical temperature is quadratic in the shear parameter $\epsilon$ but linear in the volumetric deformation $\delta$. This is a direct consequence of isotropy.

\subsubsection*{Elastic moduli}
A second and important consequence which arises from the supersolid effective action \eqref{eqsupersolid} is that the elastic moduli of the original solid are corrected by the presence of a superfluid condensate $\langle \Psi \rangle \neq 0$. More precisely, the linear elastic tensor becomes:
\begin{equation}
    \hat{C}_{\alpha\beta\gamma\delta} = C_{\alpha\beta\gamma\delta} + a_2(T)_{\alpha\beta\gamma\delta} |\Psi|^2 + b_2(T)_{\alpha\beta\gamma\delta} |\Psi|^4.
\end{equation}
In the isotropic case, using \eqref{eq:isotropy}, this simply reduces to:
\begin{equation}\label{eq:moduli}
    \begin{cases}
    \hat{\mathcal{K}} = \mathcal{K} + a_2^{(\mathcal{K})}(T) |\Psi|^2 + b_2^{(\mathcal{K})}(T) |\Psi|^4 \\
    \hat{\mathcal{G}} = \mathcal{G} + a_2^{(\mathcal{G})}(T) |\Psi|^2 + b_2^{(\mathcal{G})}(T) |\Psi|^4
    \end{cases} \, .
\end{equation}
Let us elaborate more about the correction to the shear modulus. The ground state in the broken homogeneous phase which arises from the effective action in Eq.\eqref{eqsupersolid} is defined as:
\begin{equation}
    |\Psi|^2 (T) = - \frac{a(T)}{2\, b(T)}\,,
\end{equation}
as it follows from free energy's minimization. Substituting this expression in \eqref{eq:moduli}, and going in the vicinity of the superconducting transition (so that the quartic-in-the-condensate term can be neglected), one obtains:
\begin{equation}\label{eq:shear_deviation}
    \delta\mathcal{G} = \hat{\mathcal{G}} - \mathcal{G} = a_2^{(\mathcal{G})} |\Psi|^2= a_2^{(\mathcal{G})} \, \frac{a_0}{2\, b} \, \left|T-T_c^{(0)}\right|\,,
\end{equation}
where all the parameters $a_2^{(\mathcal{G})},\, a_0,\,b$ are evaluated at the original critical temperature $T_c^{(0)}$.
An analogous formula holds for the bulk case. Most interestingly, though, we notice that, since the fraction in \eqref{eq:shear_deviation} is positive, it follows that the corrections \eqref{eq:shear_tc_corr} and \eqref{eq:shear_deviation} must have opposite sign:
\begin{equation}\label{eq:opposite_signs}
    \delta \mathcal{G} \cdot \delta T_c^{(\epsilon)} < 0.
\end{equation}

As a conclusive remark, we would like to stress that the GL model presented in this Section does not provide any of the phenomenological coefficients, which should be derived by other means.  In the remainder of this work, we will see how the adoption of holographic methods, instead, gives us access to those coefficients, thus completing in a sense the GL model.

\section{The holographic model} \label{section:model}
We consider the following four dimensional gravitational action (in natural units $\hbar = c = 1$):
\begin{equation}\label{eq:full_action}
    \mathcal{S}_{grav} = \int d^{4}x \sqrt{-g} \left\{ \frac{1}{2 \kappa^2} \left( R - 2 \Lambda - 2 m^2 V(X) \right)  -  \frac{1}{4 e_0^2} F^2 -|D\psi|^2 - M^2 |\psi|^2 \right\},
\end{equation}
which combines the original holographic superconductor model of \cite{Hartnoll_2008} with the generalized axions model of \cite{baggioli2015electron}. Here, $\Lambda$ is the AdS$_4$ cosmological constant (so that $\Lambda=-3/L^2$ with $L$ being the AdS length), $F=dA$ is the field strength of a $U(1)$ bulk gauge field $A_\mu$, $\psi$ is a complex scalar field charged under the $U(1)$ symmetry and $D_\mu=\partial_\mu-i q A_\mu$ is the standard covariant derivative. The equations of motion of the system allow to set the phase of the complex scalar $\psi$ to zero, at least at the level of the (homogeneous) background. Following dimensional analysis, $V(X)$ is a generic, dimensionless scalar potential defined in terms of the scalar variable $X\equiv \frac{1}{2} \partial_\mu \phi^I \partial ^\mu \phi^I$ with $\phi^I$, $I \in \{x,y\}$,  being a set of two massless real scalar fields (the axion fields). The function $V(X)$ is not entirely arbitrary as it must obey specific consistency requirements, derived in \cite{baggioli2015electron}. Notice that by setting $m=0$, one recovers the action of \cite{Hartnoll_2008} and the standard holographic superfluid model, which we can then consider as a limiting case of our model. 
We adopt Poincaré coordinates: $\left\{t,x,y\right\}$ for the boundary temporal and spatial coordinates, and $u$ for the radial bulk one. The UV boundary is placed at $u = 0$ while the event horizon of the black brane solution at $u=u_h$.

The action \eqref{eq:full_action} is invariant under global shifts in the axion fields: $\phi^I \rightarrow \phi^I + a^I$, with $a^I$ a constant two-dimensional vector. Because of this property, their equation of motion:
\begin{equation}
    \partial_\mu \left(\sqrt{-g} \, \frac{\partial V}{\partial X} \, g^{\mu\nu} \partial_\nu \phi^I\right)  = 0,
\end{equation}
is trivially solved by a background configuration which is linear in the spatial coordinates of the boundary $x^i$ and taken for convenience as:
\begin{equation}\label{eq:def_ansatz}
    \phi^I = {\Theta^I}_j \, \frac{x^j}{L_0},
\end{equation}
where ${\Theta^I}_j$ is a constant, dimensionless, rank 2 matrix, which we will identify, in the following Subsection \ref{subsection:holo_deformations}, with the inverse of the deformation gradient tensor (about the physical significance of this quantity, see Appendix \ref{app1}), and $L_0$ is a characteristic lengthscale necessary to have the matrix $\Theta$ dimensionless. All in all, the axion fields $\phi^I$ are dimensionless, as expected.  

We will later identify the reference configuration with ${\Theta^I}_j\equiv {\delta^I}_j$. Therefore, $L_0$ can be thought to be related to the distance between the atomic positions in the reference configuration. In order for this holographic model to be dual to a field theory exhibiting \textit{spontaneous} translational symmetry breaking, we will need to require the background solution $\phi_{ref}^I=x^I/L_0$ to correspond to the expectation value of the scalar operators dual to the bulk axion fields. This is obtained by placing further constraints on the $V(X)$ function, as shown later (see \cite{Baggioli:2021xuv} for more details).\medskip

Following \cite{baggioli2020black}, we parameterize the constant matrix appearing in \eqref{eq:def_ansatz} as:
\begin{equation}\label{bla}
    \Theta = \alpha
     \begin{pmatrix}
        \cosh{\left(\Omega/2\right)} &  \sinh{\left(\Omega/2\right)} \\
         \sinh{\left(\Omega/2\right)} &  \cosh{\left(\Omega/2\right)}
    \end{pmatrix}\,.
\end{equation}
We also take the following ansatz for the rest of the bulk fields:
\begin{equation}\label{eq:rest_ansatz}
\begin{gathered}
    ds^2 = \frac{L^2}{u^2} \left(- f(u)e^{-\chi(u)} dt^2 + \frac{du^2}{f(u)} + \gamma_{ij}(u) dx^i dx^j \right), \\[1pt]
    \gamma(u) =
    \begin{pmatrix}
    \cosh h(u) & \sinh h(u) \\
    \sinh h(u) & \cosh  h(u)
    \end{pmatrix}, \qquad A = A_t(u) dt, \qquad \psi = \psi(u),
\end{gathered}
\end{equation}
with $\gamma(u)$ obeying the property $\det(\gamma)=1$. We restrict to black-brane solutions where $f(u)=0$ at $u=u_h$. The bulk equations of motion which follow are presented in Appendix \ref{app2} together with a more detailed analysis of the symmetries of our system and their meaning. Such analysis allows us to set the AdS length $L$ to unity ($L=1$); we then make a choice for the other parameters appearing in the action \eqref{eq:full_action}, namely: $\kappa/L=1/\sqrt{2}$, $e_0=1$, $m L = 1$, $q=3$ and $(M L)^2 = -2$. This choice sets the UV asymptotics of the charged scalar field $\psi$ to be:
\begin{equation}
    \psi \simeq \psi^{(1)} u +\psi^{(2)} u^2 +\dots \,. \label{eq:vevs}
\end{equation}
Assuming a standard quantization scheme, the leading term in this expansion, $\psi^{(1)}$, corresponds to the source of the dual operator $\mathcal{O}$ and the subleading one, $\psi^{(2)}$, to its expectation value $\langle \mathcal{O}\rangle$. Background solutions in the broken phase ($T < T_c$) are found by solving numerically the bulk equations of motions. Details are reported in Appendix \ref{app2}. As a benchmark model, we will consider a potential of the form:
\begin{equation}
    V(X)\,= \,X^N\,,\qquad \text{with}\qquad N>5/2\,.
\end{equation}
 The choice of the exponent guarantees that (in the standard quantization scheme) the operators dual to the axion fields $\phi^I$ break the translational invariance of the dual field theory spontaneously, making such a system a \textit{solid}. For a complete description and understanding of this mechanism we refer to \cite{Alberte:2017oqx,Ammon:2019wci,Baggioli:2021xuv}. For the remainder of this work, we set $N=3$. In making such a choice, we focus on a field theory that has no canonical kinetic term. This may raise a question on whether the time evolution of the system is non-unitary; however, we stress that the theory at hand \textit{does not} have any higher derivative kinetic term, rather a higher power of the usual two-derivative one. As such, no unitarity problem arises; instead, the unusual feature of theories of this form is that they make sense only around non-trivial vacuums (and not around the trivial one $\phi^I=0$ where the model becomes strongly coupled). We refer to \cite{Alberte:2017oqx} for a more detailed discussion as for why that is the case and which are the physical implications.

 Finally we can define the the temperature $T$, the charge density $\rho$ and the chemical potential $\mu$ of the dual field theory using the standard AdS-CFT dictionary \cite{zaanen2015holographic,Baggioli:2019rrs}:
\begin{align}
   & T\,=\,-\frac{f'(u_h)}{4\pi}\,e^{-\chi(u_h)/2}\,,\qquad \mu\,=\,A_t(0)\,,\qquad \rho\,=\,-A_t'(0)\,,
\end{align}
where primes indicate differentiation with respect to the radial coordinate $u$.

To summarize, any state of the boundary theory can be then parameterized by the value of the three following dimensionless parameters:
\begin{equation}\label{eq:holo_labels}
    \{T L_\alpha, \, \rho L_\alpha^2, \, \Omega\},
\end{equation}
where we have defined $L_\alpha=L_0/\alpha$. Unsheared configurations corresponds to $\Omega=0$.\\

\subsection{On the definition of mechanical deformations in holography}\label{subsection:holo_deformations}

Before proceeding to the results of our work, let us briefly describe how the holographic model at hand naturally includes mechanical deformations. Our notations follow closely the ones adopted in the EFT framework of \cite{Alberte:2018doe,Pan:2021cux}. Importantly, we will assume the deformations to be finite, or even large, and we will therefore consider the full framework of nonlinear elasticity. For a brief summary of nonlinear elasticity theory we refer to Appendix \ref{app1}; a more in depth treatment can be found in \cite{ZAMM:ZAMM19850650903}. \medskip

The starting point of our analysis is the definition of the Lagrangian strain tensor (see \cite{Armas:2019sbe}). Let us consider a medium in a space equipped with a background flat metric $g_{\mu\nu}=\eta_{\mu\nu}$. The distance between points in such a medium can be computed by using the \textit{crystal metric} defined as:
\begin{equation}
    ds^2_{cr} = h_{IJ} \left( e^I_\mu \,dx^\mu \right) \left( e^I_\nu \,dx^\nu \right),
\end{equation}
where $\mu$, $\nu$ are indices running on spacetime coordinates, and $e^I_\mu(x)$ are a set of one-form fields, with $I \in \{1,...,d\}$. Deformations are then defined with respect to an initial reference configuration:
\begin{equation}
    ds^2_{ref} = h^{ref}_{IJ} \left( e^I_\mu dx^\mu \right) \left( e^I_\nu dx^\nu \right)\,.
\end{equation}
 For practical purposes, we will later make the convenient choice $h^{ref}_{IJ} = \delta_{IJ}$.

Once a reference configuration is selected, it is straightforward to define the Lagrangian strain tensor $E_{IJ}$ \cite{chaikin_lubensky_1995} as:
\begin{equation}\label{eq:jain_armas_strain_tensor}
    ds^2_{cr} - ds^2_{ref} = 2 E_{IJ} \left( e^I_\mu dx^\mu \right) \left( e^I_\nu dx^\nu \right) \quad \Longrightarrow \quad  E_{IJ} = \frac{1}{2}\left(h_{IJ} - h^{ref}_{IJ} \right).
\end{equation}
In absence of defects, the medium can be smoothly parameterized by mean of the \textit{crystal fields} $\varphi^I$ which serve as the \textit{material coordinates} (see Appendix \ref{app1} for more details). Then, the one forms $e^I_\mu(x)$ can be written as:
\begin{equation}
    e^I_\mu = \partial_\mu \varphi^I\,.
\end{equation}
The material coordinates $\varphi^I$ are related to the axion fields $\phi^I$ appearing in our holographic model via the simple identification:
\begin{equation}\label{eq:displacement_def}
    \varphi^I = L_0 \,\phi^I.
\end{equation}
Making such a combination is necessary to take into account the fact that the axion fields $\phi^I$ are dimensionless Goldstone phases, and cannot be directly identified with spatial coordinates. As a direct consequence of Eq.\eqref{eq:displacement_def}, the reference configuration\footnote{Within the holographic picture, this configuration does not minimize the free energy and possesses a residual background stress labelled \textit{crystal pressure}  in \cite{Armas:2019sbe}. Nevertheless, one can still consider linearized perturbations around such a state, and show that their dynamics is perfectly consistent with the hydrodynamics expectations \cite{Ammon:2020xyv,Baggioli:2020edn}.} is then given by:
\begin{equation}\label{refc}
    \varphi_{ref}^I = {\delta^I}_j \, x^j.
\end{equation}

Using the previous results, the inverse of the crystal metric can easily be written in terms of the crystal fields as:
\begin{equation}
    h^{IJ} = \left(h_{IJ}\right)^{-1} = g^{\mu\nu} \partial_\mu \varphi^I \partial_\nu \varphi^J.
\end{equation}
This provides us with everything that we need to compute the strain tensor \eqref{eq:jain_armas_strain_tensor} in our holographic model. By using \eqref{eq:displacement_def} along with \eqref{eq:def_ansatz} in the above expression, we find:
\begin{equation}
    h^{IJ} = \delta^{ij} {\Theta^I}_i {\Theta^J}_j \quad \Longleftrightarrow \quad h^{-1} = \Theta^T \Theta\,.
\end{equation}
Using \eqref{eq:jain_armas_strain_tensor}, we then obtain:
\begin{equation}
    E = \frac{1}{2} \left( (\Theta^{-1})^T \Theta^{-1} - \mathbb{I} \right)\,.
\end{equation}
 By comparing this expression with the standard definition of the Lagrangian strain tensor in nonlinear elasticity theory, Eq.\eqref{efull}, we find the important result:
\begin{equation}
    \Xi = \Theta^{-1},
\end{equation}
where $\Xi$ is the deformation gradient tensor. As expected, the inverse of the deformation gradient tensor is given by the Jacobian matrix of the material coordinates with respect to the spacetime ones. \medskip

In the rest of the manuscript, we will consider two separate cases: (I) a pure volumetric, shape-preserving deformation, and (II) a purely deviatoric strain, which preserves the volume of any subregion of the medium throughout the deformation process.

The first scenario is realized by setting $\Omega=0$ and identifying:
\begin{equation}\label{qq}
   \delta=\frac{\Delta V}{V}\,=\,\mathrm{Tr}\left[E_{ij}\right]\,=\,\mathrm{det} \,\Xi\,=\left(\frac{1}{\alpha^2} - 1 \right)\,.
\end{equation}
From this expression, a dilatation, $\delta>0$, corresponds to the choice $\alpha<1$ and a compression with $\alpha>1$. This is consistent with the fact that $L_\alpha$ is related to the lattice size of the system; the case $\alpha=1$ corresponds to absence of mechanical deformations (with respect to the reference configuration \eqref{refc}). Finally, a linear isotropic deformation corresponds to an infinitesimal shift $\alpha \rightarrow \alpha+\Delta \alpha$, under which:
\begin{equation}
    \delta=\frac{\Delta V}{V}= - 2 \,\Delta \alpha,
\end{equation}
where we have fixed $\alpha=1$. This minus sign is consistent with previous arguments \cite{baggioli2015electron,Baggioli:2019elg,Armas:2020bmo} and with the physical picture that $\Delta \alpha>0$ implies a reduction of the solid length-scale $L_\alpha$. \medskip

The second scenario, that of a deviatoric strain that causes no changes in the volume of any subregion of the medium, can be realized by setting $\alpha=1$ and by noticing that the shear strain is parameterized by:
\begin{equation}\label{eq:def_epsilon}
    \Xi_{xy} \equiv \frac{1}{2}\,\epsilon = - \sinh{\left(\frac{\Omega}{2}\right)}\,.
\end{equation}
As anticipated, at nonlinear level, we define $\epsilon$ through the off-diagonal component of the deformation gradient tensor, which no longer exactly coincides with the off-diagonal component of the nonlinear strain tensor $E_{xy}$.\footnote{For completeness, we report that:
\begin{equation}
    E_{xy}=-\frac{\sinh{\Omega}}{2},
\end{equation}
which is slightly different from $\Xi_{xy}$ in Eq.\eqref{eq:def_epsilon} in the nonlinear regime. In the linear regime, the two coincide:
\begin{equation}
    \Xi^{lin}_{xy}=E^{lin}_{xy}=-\frac{\Omega}{2}\,.
\end{equation}}

Finally, the minus sign appearing in \eqref{eq:def_epsilon} can be dropped if we extend the $SO(2)$ symmetry to the full orthogonal group $O(2)$. It can indeed easily be seen that, given the rank 2 representation of the reflection with respect to the $x$ axis, which we denote by $\sigma_x$:
\begin{equation}
    \sigma_x = \begin{pmatrix}
        1 & 0 \\
        0 & -1
  \end{pmatrix} \, \Rightarrow\, \begin{pmatrix}
        \cosh{\left(\Omega/2\right)} &  - \sinh{\left(\Omega/2\right)} \\
       - \sinh{\left(\Omega/2\right)} &  \cosh{\left(\Omega/2\right)}
    \end{pmatrix} = \sigma_x \begin{pmatrix}
   \cosh{\left(\Omega/2\right)} &  \sinh{\left(\Omega/2\right)} \\
         \sinh{\left(\Omega/2\right)} &  \cosh{\left(\Omega/2\right)}
    \end{pmatrix} \sigma_x,
\end{equation}
so that the minus sign appearing in \eqref{eq:def_epsilon} can be reabsorbed by the action of the reflection $\sigma_x$. Because of this reason, the minus sign will be dropped in the rest of the manuscript.

In summary, to compare the effective description presented in Section \ref{effectivesection} with the results from holography, we just need to use the simple identifications:
\begin{equation}
    \delta=\left(\frac{1}{\alpha^2} - 1 \right)\,,\qquad \epsilon=2 \sinh\left(\frac{\Omega}{2}\right)\,,\label{dd1}
\end{equation}
where now the parameters $\delta,\,\epsilon$ are taken at full nonlinear level. The linear regime can be re-obtained by taking the limits $\alpha,\, \Omega \ll 1$.\\
Let us emphasize that since the reference configuration defined in Eq.\eqref{refc} does not minimize the free energy, all configurations $\phi^I=\alpha x^I/L_0$ could be taken as legitimate reference choices. Given an initial state with $\alpha=\alpha_i$ and a final state with $\alpha=\alpha_f$, the parameter $\delta$ encoding a purely volumetric deformation can be more generally defined as:
\begin{equation}\label{law}
    \delta=\left(\frac{1}{\alpha_f^2}-\frac{1}{\alpha_i^2}\right),
\end{equation}
which boils down to the previous definition \eqref{dd1} when $\alpha_i=1$ and $\alpha_f=\alpha$, as in our choice of reference configuration.\medskip

\section{Results}
\label{sec:results}
In this section, we outline the results obtained via the holographic model \eqref{eq:full_action} and we compare them with the effective theory of Section \ref{effectivesection}.

\subsection{The phase diagram of the unsheared solid}

As a first step of our study, we build the phase diagram of the theory described by \eqref{eq:full_action} for unsheared ($\epsilon=0$) configurations. More precisely, we compute the dimensionless critical temperature $T_c L_\alpha$, at which the global $U(1)$ symmetry is spontaneously broken, in terms of the dimensionless charge density $\rho L_\alpha^2$. Unlike the global $U(1)$, translational invariance is always broken, indicating that our system in the normal phase ($T>T_c$) is a solid phase with finite shear rigidity. From a physical perspective, we can therefore identify the broken phase with the supersolid phase in which both translations and the $U(1)$ symmetry are spontaneously broken.

\begin{figure}
    \centering
    \includegraphics[trim=25 0 0 0,clip,width=0.45 \linewidth]{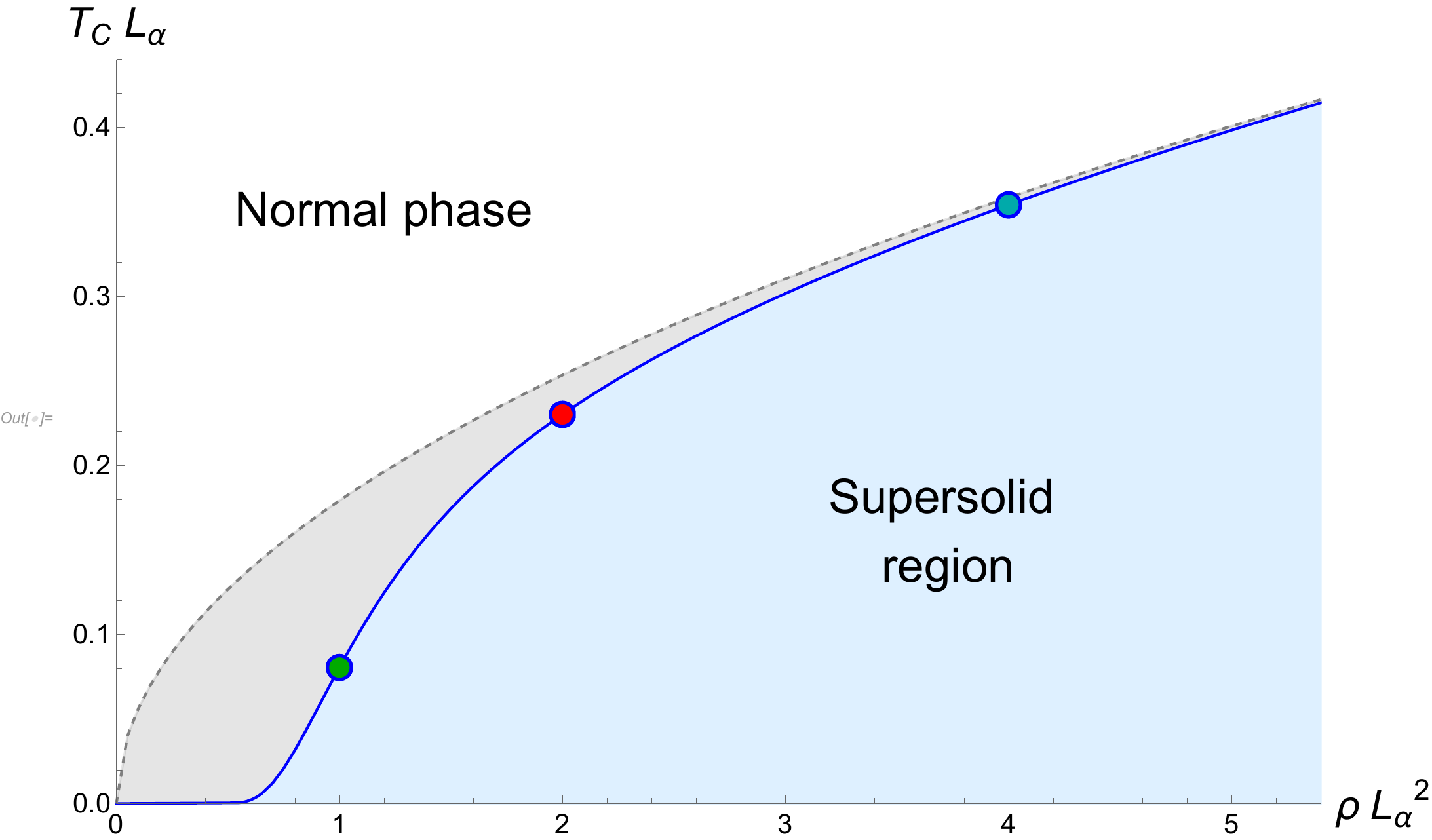} \qquad 
    \includegraphics[trim=25 0 0 0,clip,width=0.45 \linewidth]{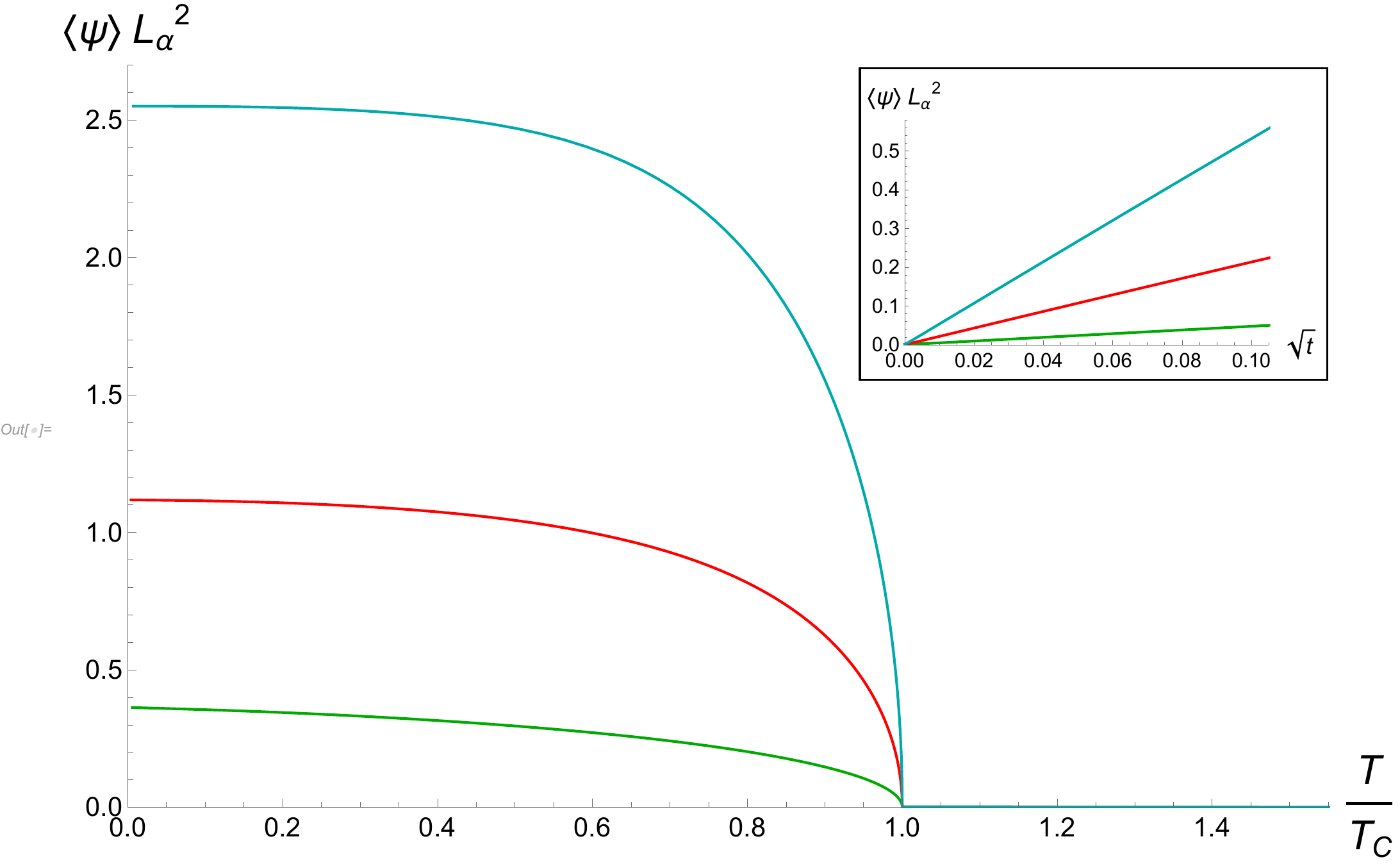}
    \caption{\textbf{Left:} The critical temperature as a function of the charge density for unsheared configurations ($\epsilon = 0$). Colored circles indicate the following values: green ($\rho L_\alpha^2 = 1$), red ($\rho L_\alpha^2 = 2$) and cyan ($\rho L_\alpha^2 = 4$). The grey area is the set of normal states that would be superfluid in absence of elasticity (cf. \cite{Hartnoll_2008}). \textbf{Right:} The normalized condensate as a function of the temperature, normalised by its critical value, for the highlighted values of $\rho L_\alpha^2$. \textbf{Inset:} A zoom close to the transition, where the condensate scales like $|\Psi| \sim \sqrt{t}$, with $t \equiv |T-T_c|/T_c$ being the reduced temperature.}
    \label{fig:1}
\end{figure}
After these few premises, we present, in the left panel of Fig.\ref{fig:1}, the phase diagram for the unsheared configurations of our system. The blue shaded area below the solid curve indicates the supersolid phase. The dashed grey line, instead, is the critical temperature in absence of any solid component, corresponding to the original model of \cite{Hartnoll_2008}, with the grey area indicating all those states that would break the $U(1)$-symmetry if the system were not a solid. The grey shaded curve follows a square-root law as a consequence of the fact that, in the model of \cite{Hartnoll_2008}, the dimensionless critical temperature cannot be tuned, $T_c \, / \sqrt{\rho} = const$. This does not happen in presence of translational symmetry breaking, due to which a second dimensional scale -- the lattice spacing $L_\alpha$ -- appears.\medskip

We observe that higher values for the adimensional charge density correspond, in general, to a higher critical temperature $T_c$. As expected, we also find that the effects of a solid component, $L_\alpha \neq 0$, become more important in the small charge density limit. On the contrary, for large $\rho L_\alpha^2$, the bulk axions fields can be neglected and the system effectively behaves as in the original superfluid model of \cite{Hartnoll_2008}. \medskip

In the right panel of Fig.\ref{fig:1} we show the superfluid condensate $\langle \Psi \rangle$ as a function of the temperature for three different values of parameters (highlighted in the left panel); the larger the value of the adimensional charge density, the larger the (adimensional) superfluid condensate. We also find, in accordance with \cite{Hartnoll_2008}, that the supersolid transition is a mean field one, as emphasized in the inset where the typical square-root in the reduced temperature $t$ behavior of the condensate is observed:
\begin{equation}
    \langle \Psi \rangle \sim \Psi_0\, t\,,\qquad t\equiv \frac{T-T_c}{T_c},
\end{equation}
with the slope $\Psi_0$ too increasing with the charge density.

\subsection{Response to shear deformations across the supersolid phase transition}

Among the reasons behind holography's success is the relative ease with which it allows to compute Green functions. In this section, we take advantage of this feature to characterise how the holographic supersolid system reacts mechanically under an infinitesimal (within linear response theory) shear deformation. We will use the holographic prescription (see for example \cite{Baggioli:2019rrs}) to calculate numerically the retarded Green function of the off-diagonal component of the stress tensor $T^{xy}$, which at zero momentum and small frequency is\footnote{Here, we have already subtracted the standard contribution $\delta T^{xy}\sim p \,h_{xy}$, with $p$ being the thermodynamic pressure, which appears in the holographic computation as a counterterm from the holographic renormalization. See for example \cite{Burikham:2016roo}.}:
\begin{equation}\label{eq:shear_green_fct}
    \mathcal{G}^R_{T^{xy}T^{xy}}(k=0,\omega) \simeq \mathcal{G} - i \omega \eta + O(\omega^2)\,,
\end{equation}
where $\mathcal{G}$ is the shear modulus, and $\eta$ the corresponding viscosity. We will take advantage of this expansion to write the corresponding Kubo formulas:
\begin{equation}
    \mathcal{G} = \lim_{\omega\to0}
    \mathrm{Re}\,\mathcal{G}^R_{T^{xy}T^{xy}}(0,\omega), \quad \eta = \lim_{\omega\to0}
    - \frac{\mathrm{Im}\,\mathcal{G}^R_{T^{xy}T^{xy}}(0,\omega)}{\omega}\,.
\end{equation}
From a holographic point of view, the Green function \eqref{eq:shear_green_fct} is computed by introducing a spatial, transverse metric perturbation of the metric, $\delta g_{xy}(u,t) = \zeta(u)e^{-i \omega t}$. One then needs to solve numerically (with appropriate boundary conditions, outlined e.g. in \cite{Alberte:2016xja}) its finite-frequency equation of motion:
\begin{equation}
    \zeta'' + \left(\frac{f'}{f}-\frac{\chi'}{2}-\frac{2}{u}\right) \zeta' + \left(\frac{\omega ^2 e^{\chi}}{f^2}-\frac{L^2 m^2 \partial_u V(X)}{u f}\right) \zeta = 0\,.
\end{equation}
Near the UV boundary:
\begin{equation}
    \zeta(u) \simeq \zeta_0 \, (1 + ...) + \zeta_3 \, u^3 \, (1 + ...)\, ,
\end{equation}
so that we can read off:
\begin{equation}
    \mathcal{G}^R_{T^{xy}T^{xy}}(\omega) = - \frac{3}{2}\frac{\zeta_3}{\zeta_0}\,.
\end{equation}
Albeit the theory outlined in Section \ref{effectivesection} provides predictions only for the real part of the Green function \eqref{eq:shear_green_fct} (the shear modulus), we expand the scope of this section to include results for the shear viscosity $\eta$ as well. We do this for two reasons: on one hand, holography provides such results with no additional effort, making our analysis more complete; on the other hand, and most importantly, because we are interested in studying the Kovtun-Son-Starinets (KSS) bound violation in the system at hand. Such violation is expected on the general ground that the metric perturbation is massive -- as outlined in \cite{Hartnoll_bumpyBH}. KSS bound violations in a holographic axion models are widely reported in the literature \cite{Hartnoll_bumpyBH,Burikham:2016roo,Alberte:2016xja,Ling:2016ien} but definitely not totally understood from a physical perspective \cite{Baggioli:2020ljz}. The addition of a superfluid transition on top of these affects has not been the subject of a study yet and it is therefore interesting to be pursued.\\

For completeness, we show the response of our holographic model to shear deformations at finite charge density in Appendix \ref{app3}. From there, one can see that the presence of charge density diminishes the rigidity of the system (the shear modulus) and increases the value of the $\eta/s$ ratio towards its maximal value corresponding to the KSS bound. We also find that the effects of the charge density are more relevant at small temperature. Finally, we briefly show how the charge density affects the nonlinear elastic response. Also in that case we observe a decrease in the stress which is more dramatic at small shear strains.

\subsubsection{Shear modulus}\label{subsub:shear_mod}

From a macroscopic point of view, the most important physical difference between a fluid and a solid is the presence, in the latter, of a finite static shear modulus, which defines the shear rigidity of the system; such a quantity vanishes in fluid phases of matter.\footnote{See \cite{Baggioli:2019jcm,Baggioli:2021ntj} for more details about the subtle distinction between solids and liquids.} In a wide class of holographic axion models, to which the one considered in this work belongs, the shear modulus has already been computed (in \cite{Alberte:2015isw,Alberte:2016xja,Baggioli:2019elg,Andrade:2019zey}), and it has been directly used to match the speed of the transverse phonon excitations extracted numerically from the QNMs of the system with the hydrodynamic expectations \cite{Alberte:2017oqx}. In this work, we investigate how the additional $U(1)$ symmetry breaking affects the shear rigidity properties of a system that condenses in a supersolid phase below a given temperature.
\begin{figure}
    \centering
    \includegraphics[trim=25 0 0 0,clip,width=0.45 \linewidth]{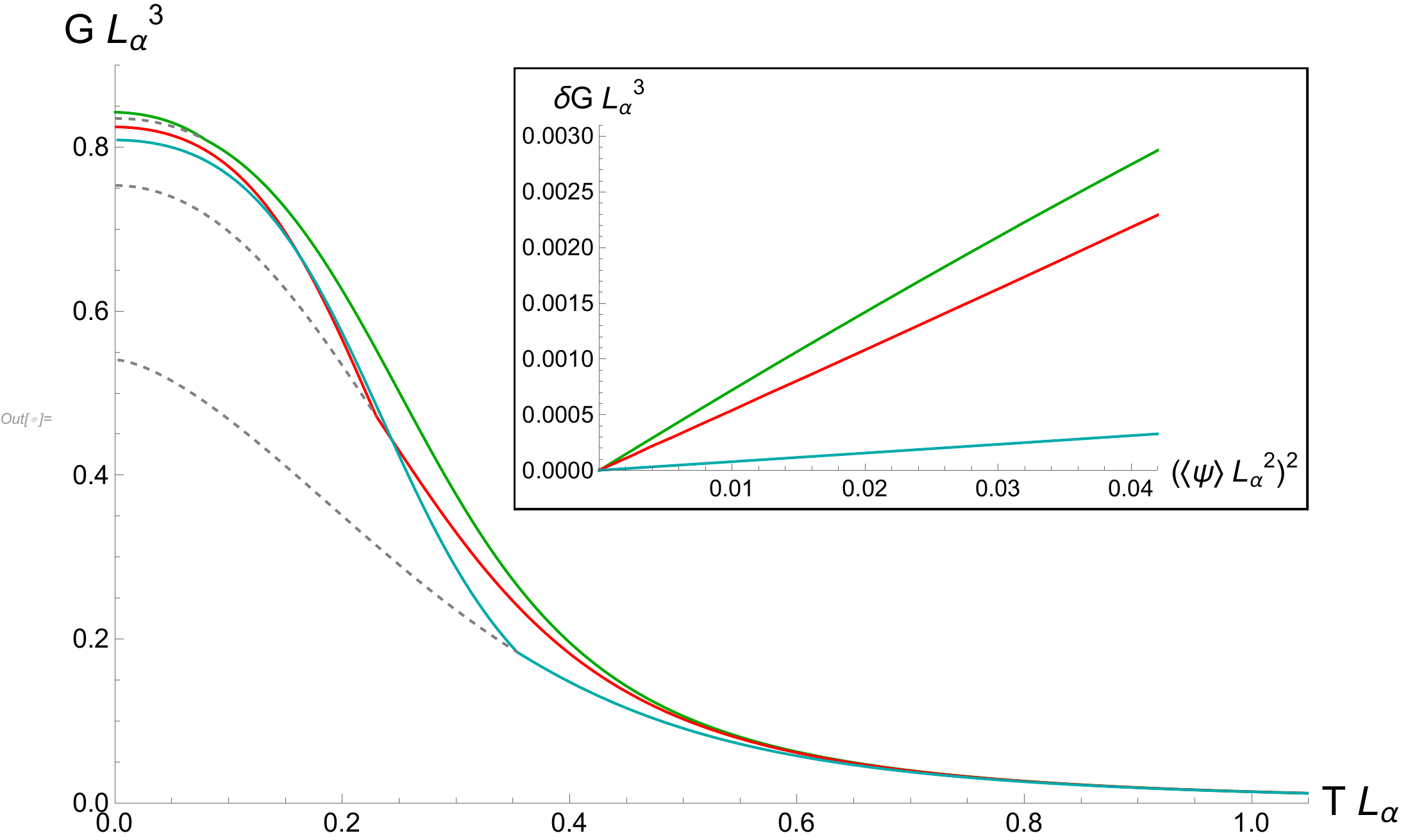} \qquad
    \includegraphics[trim=25 0 0 0,clip,width=0.45 \linewidth]{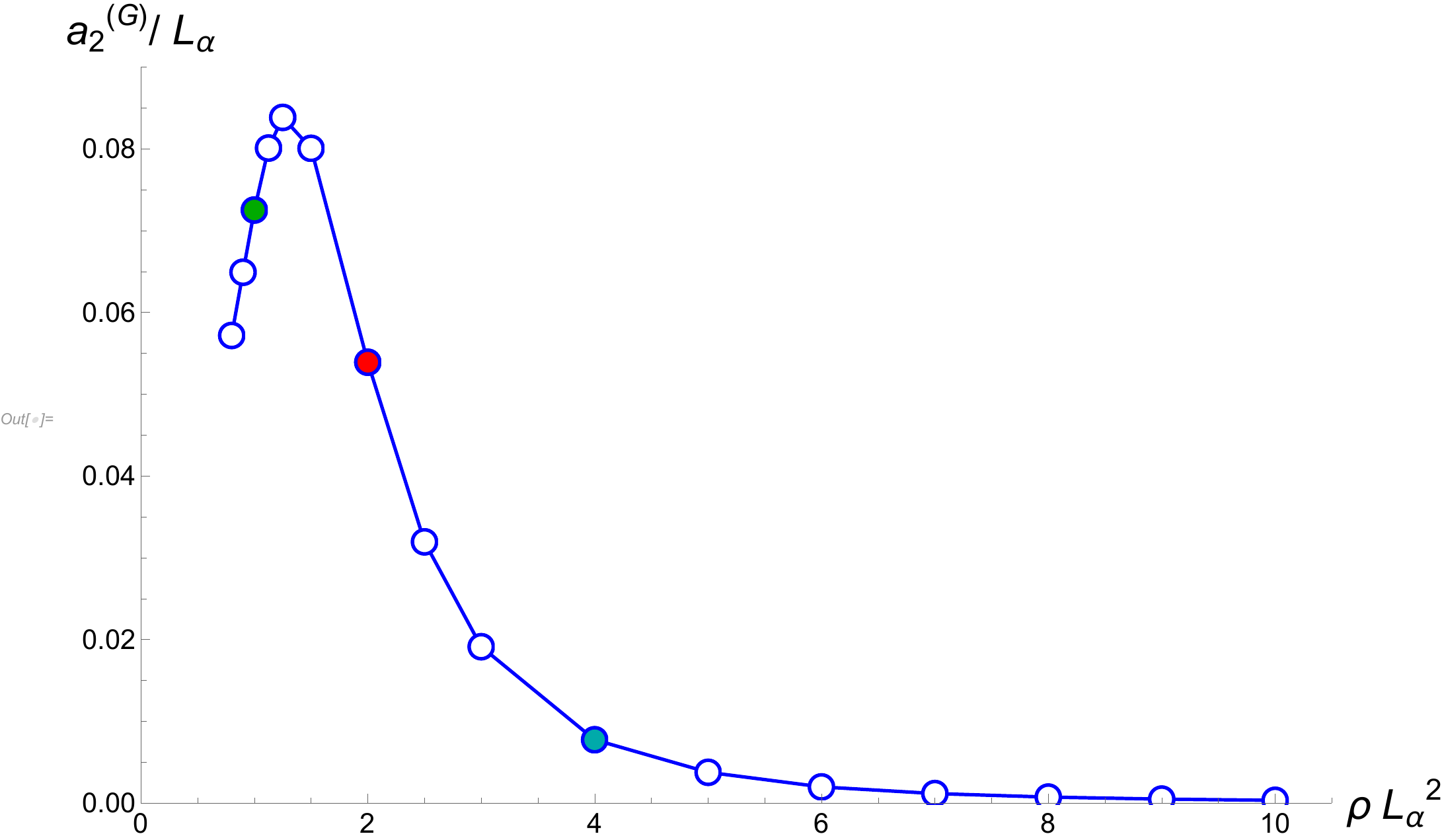}
    \caption{\textbf{Left:} Dimensionless shear modulus as a function of $T L_\alpha$. Data series are identified by their colour, see Fig. \ref{fig:1}. Dashed lines report the would-be value for the shear modulus in absence of a superfluid condensate. \textbf{Inset:} The correction to the shear modulus near the transition, as a function of $|\Psi|^2$, showing the linear behaviour predicted by \eqref{eq:shear_deviation}. \textbf{Right:} The coefficient $a_2^{(\mathcal{G})}$ appearing in \eqref{eq:shear_deviation} as a function of $\rho L_\alpha^2$.}
    \label{fig:2}
\end{figure}

In Fig.\ref{fig:2}, we plot the dimensionless shear modulus in function of the adimensional temperature, spanning both the normal and the supersolid phases. We observe a strong enhancement of the shear modulus $\mathcal{G}$ taking place in the supersolid phase: coloured solid lines indicate $\mathcal{G}$ in the full system, while dashed ones report its would-be value in a system that does not condense. Upon the transition the shear modulus is continuous, but its first derivative is not. Near the phase transition, at small values of the superfluid condensate $\langle \Psi \rangle$, we observe a linear-in-$|\Psi|^2$ increase of the shear modulus:
\begin{equation}
    \delta \mathcal{G} \sim |\Psi|^2,
\end{equation}
where $\delta \mathcal{G}$ is defined as the difference between the shear modulus in the supersolid phase and the corresponding value for the solid phase with no superfluid condensate. The inset of the left panel of Fig.\ref{fig:2} shows more clearly this feature.

Our numerical findings are in agreement with the EFT expectations of Eq. \eqref{eq:shear_deviation}, thus allowing us to derive the value of the EFT parameter denoted therein as $a_2^{(\mathcal{G})}$ for a set of values of the adimensional charge density. In the right panel of Fig.\ref{fig:2}, we show that such a parameter displays a distinct non-monotonic behaviour in function of $\rho L_\alpha^2$. More precisely, it increases for small values of $\rho L_\alpha^2$ up to a maximum $(\rho L_\alpha^2)_{max} \approx 1$, before decreasing monotonically towards zero. We find it significant that the maximum value, which signals the maximum interplay between superfluidity and elasticity, occurs exactly when the parameter $\rho L_\alpha^2$ is $\mathcal{O}(1)$. Physically that is the point at which the effects of charge density and shear ridigity are comparable and none of the them can be neglected. On the contrary, the regimes where $\rho L_\alpha^2 \ll 1$ or $\rho L_\alpha^2 \gg 1$, corresponds to situations where one of the two dimensional scales prevails on the other rendering the other irrelevant. Finally, we find that for, large charge density $\rho L_\alpha^2 \gg 1$, the effects of a small superfluid condensate on the solid behaviour (parameterized by the leading order coefficient $a_2^{(\mathcal{G})}$) are negligible. Away from the transition (for large values of the condensate), on the contrary, the effect seems to be stronger for those states. At the same time, we do expect that for small $\rho L_\alpha^2$ the effects of the superfluid condensate are vanishing as well since the latter tends to zero (see left panel of Fig.\ref{fig:1}). \medskip

Interestingly, a sudden increase of the shear modulus at the supersolid transition has been reported experimentally \cite{doi:10.1063/1.2883895,Day2007,Syshchenko_2009}.

\subsubsection{Shear viscosity}

\begin{figure}
    \centering
    \includegraphics[trim=25 0 0 0 ,clip,width=0.45 \linewidth]{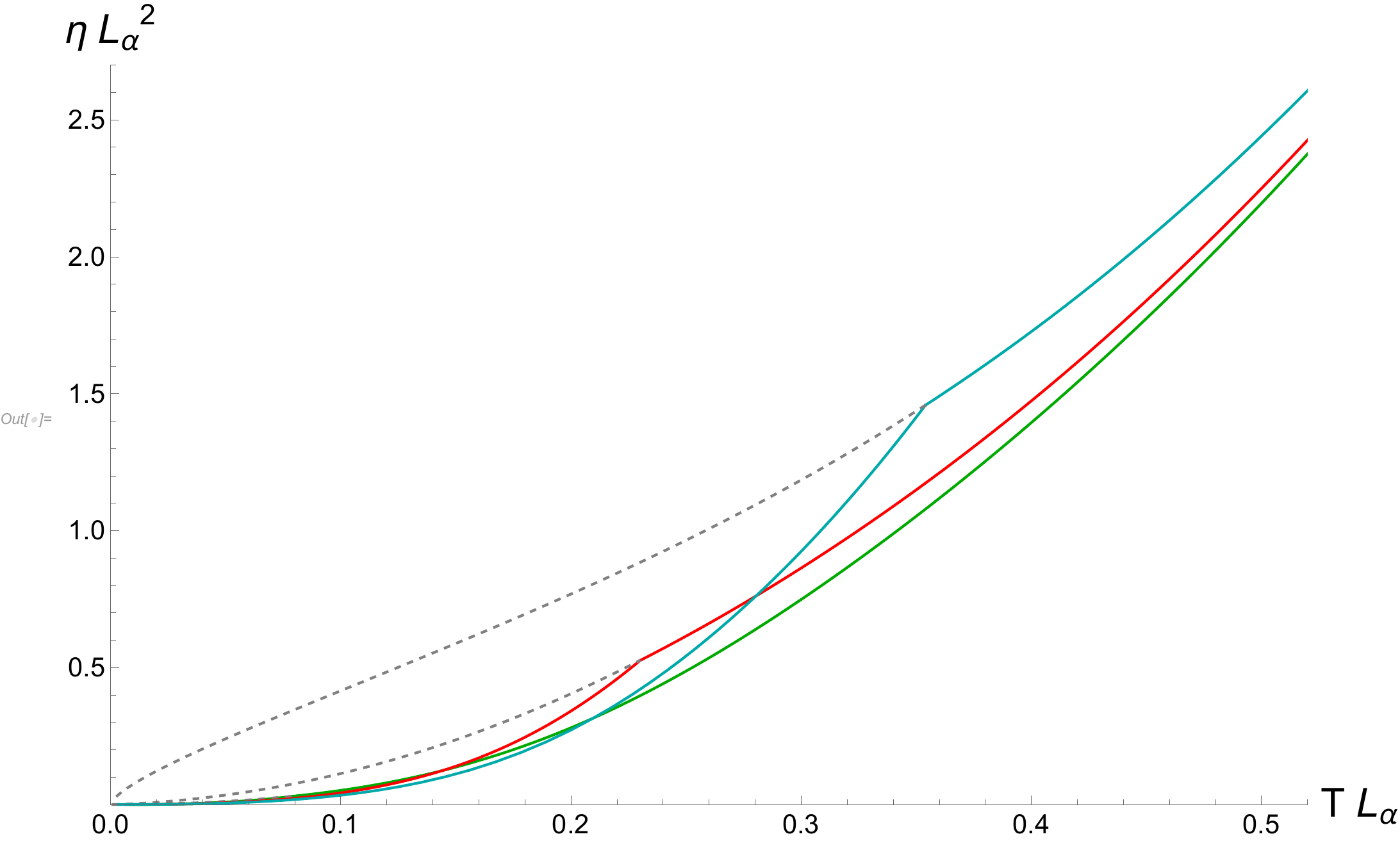} \qquad 
    \includegraphics[trim=25 0 0 0 ,clip,width=0.45 \linewidth]{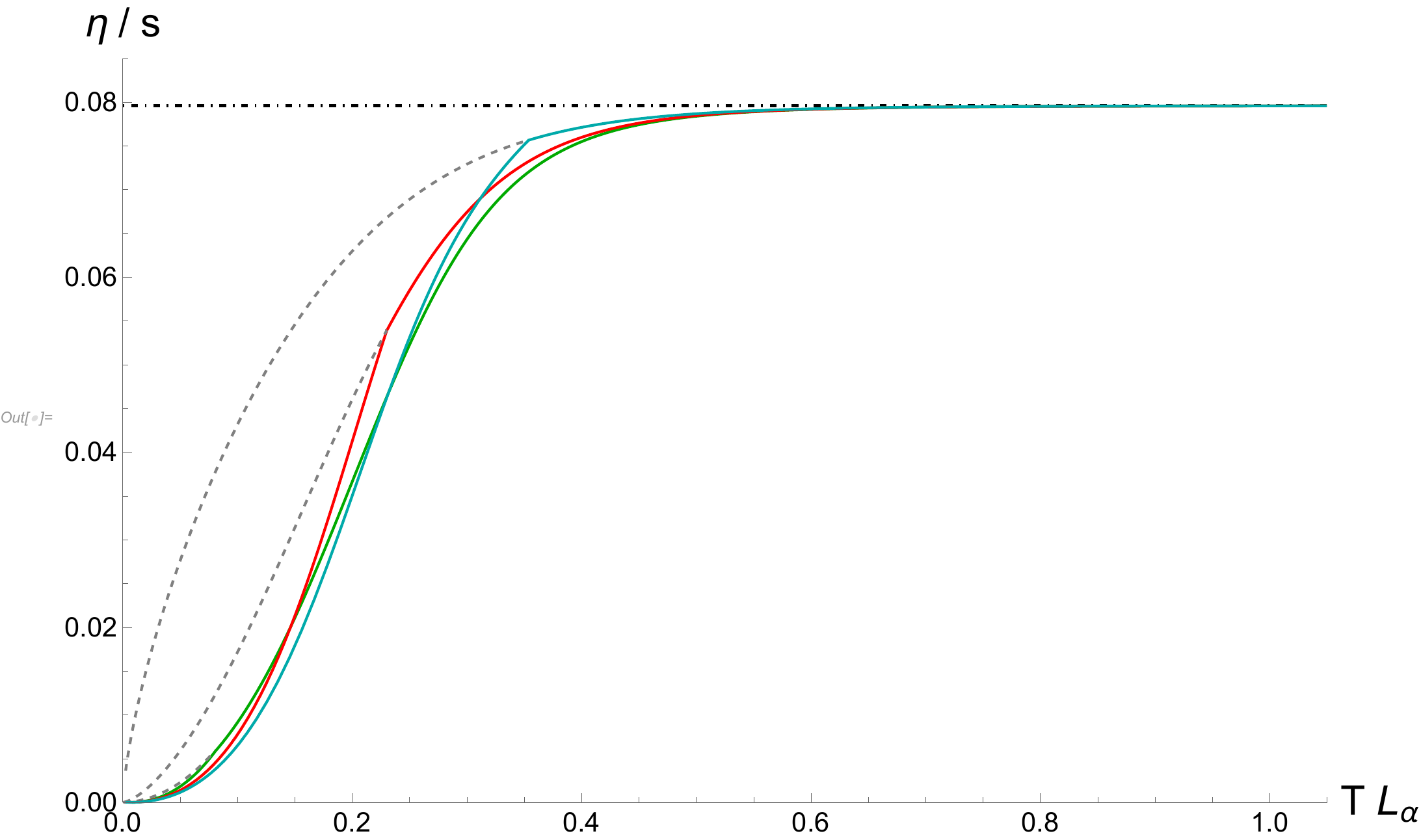}
    \caption{\textbf{Left:} Dimensionless shear viscosity as a function of $T L_\alpha$. Data series are identified by their colour, see Fig. \ref{fig:1}. Dashed lines report the would-be value for the shear viscosity in absence of a superfluid condensate. The discontinuous points correspond to the critical temperature and the onset of supersolidity. \textbf{Right:} KSS bound violation for the same data series; the dashed lines indicate would-be values in absence of condensation, while the dot-dashed one is the KSS bound: $\frac{\eta}{s} = \frac{1}{4 \pi}$.}
    \label{fig:viscosity}
\end{figure}

We now turn to the information that we can extract from the imaginary part of the Green function \eqref{eq:shear_green_fct} -- the shear viscosity $\eta$. Results coming from the numerical integration of the equation of motion for the transverse metric perturbation are reported in Fig.\ref{fig:viscosity}. The qualitative features of the adimensionalized $\eta$ at the transition temperature -- left panel -- are similar to the ones reported for the shear modulus $\mathcal{G}$. The shear viscosity appears to be continuous upon the transition, unlike its first derivative. Importantly, the correction is negative implying that the viscosity in the supersolid phase is always lower than in the solid one (with the same choice of parameters). This is intuitively in accordance with the physical idea that in the supersolid phase part of the system, i.e. the superfluid component (cf. two-fluid model), flows without friction, resulting in an overall lower viscosity.
In the right panel of Fig.\ref{fig:viscosity} we report the behavior of the viscosity-to-entropy ratio $\eta/s$. There, $s$ is the entropy density of the dual field theory extracted from the Bekenstein-Hawking entropy of the black hole in the bulk. With our notations, we have:
\begin{equation}
    s = \frac{2 \pi}{u_h^2}\,,
\end{equation}
with $u_h$ being the radial coordinate's value on the horizon. As anticipated before, the KSS bound is violated for all values of the adimensional charge density considered, at all temperatures and both in the solid and supersolid phases. This is expected because of the massive nature of gravity in the model at hand. However, we find that the condensation makes such violation even more pronounced. Interestingly, a pure holographic $s$-wave superfluid is reported to not violate the KSS bound \cite{Natsuume_shearvisc}. From this fact, we draw the conclusion that the mechanism that enhances the KSS bound violation is to be found in the gravity-mediated interaction between the condensate and the axions -- that, in our model, play the role of the phonons. From the dual field theory perspective, this interaction is rooted in the low-energy description of the supersolid state and responsible for most of the effects described in this work.

\subsection{Critical temperature under mechanical deformations}

Now we turn to investigate the supersolid critical temperature $T_c$ of the system, with the specific aim to understand how it is affected by the parameters describing the deformations of the solid.\medskip

We start by considering a purely deviatoric deformation of the system, described by the shear strain parameter $\epsilon$.

\begin{figure}
    \centering
    \includegraphics[trim=25 0 0 0 ,clip,width=0.45 \linewidth]{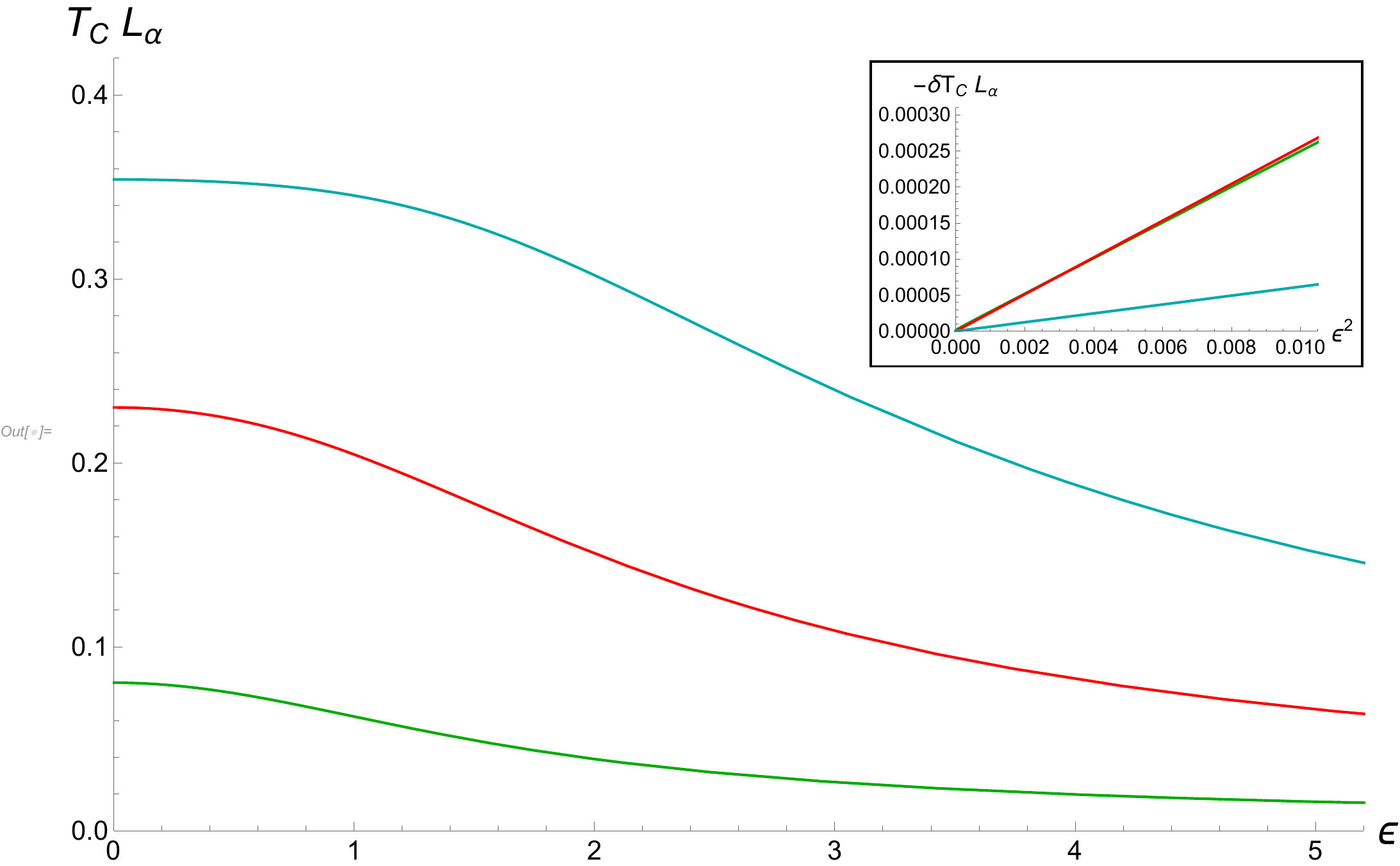} \qquad 
    \includegraphics[trim=25 0 0 0 ,clip,width=0.45 \linewidth]{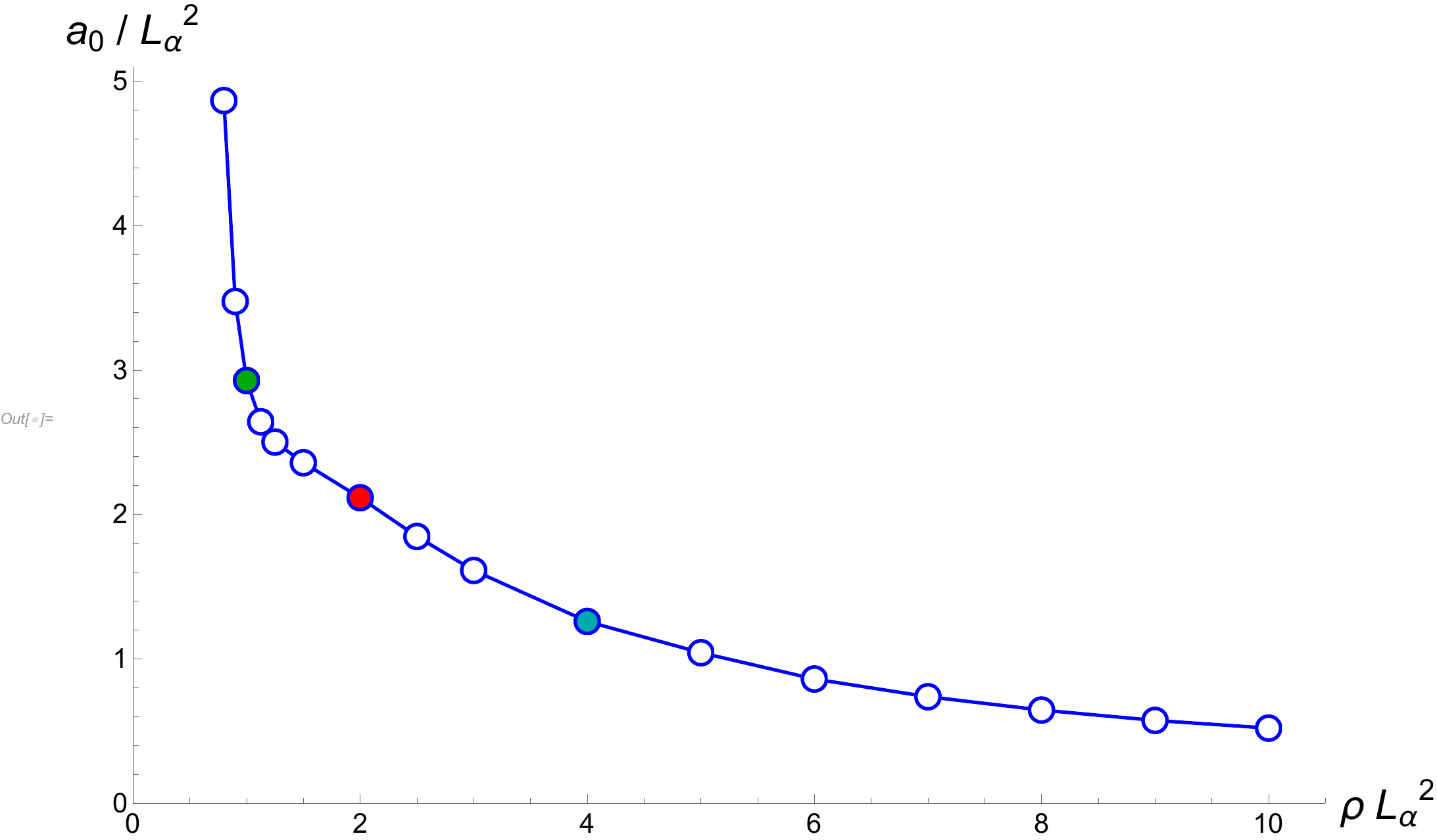}
    \caption{\textbf{Left:} The critical temperature as function of the shear parameter $\epsilon$ for the notable values of $\rho L_\alpha^2$ introduced in Fig.\ref{fig:1}. The starting ($\epsilon=0$) point of each curve should be identified with the corresponding highlighted point in Fig.\ref{fig:1}. The inset shows how small deformations induce a quadratic decrease of the critical temperature as a function of the shear parameter $\epsilon$, again in agreement with the GL theory \eqref{eq:shear_tc_corr}. \textbf{Right:} The normalized coefficient $a_0$ appearing in \eqref{eq:shear_tc_corr} as a function of the adimensional charge density.}
    \label{fig:3}
\end{figure}

In the left panel of Fig.\ref{fig:3}, we show the dimensionless supersolid critical temperature in function of the shear strain $\epsilon$ for some values of the dimensionless charge density. The value of $\rho L_\alpha^2$ does not seem to be affecting the qualitative behaviour of the curves, which all start horizontally with a negative concavity, then later have an inflection point while continuing to decrease monotonically. No curve is observed to reach a vanishing critical temperature for a finite value of the shear strain parameter $\epsilon$; it is possible indeed to check that the zero temperature state violates the near horizon BF bound, so that the supersolid phase is expected to survive up to arbitrarily large strains and arbitrarly low temperatures. Let us notice that the larger the charge density the larger the shear strain needed to affect significantly the critical temperature.

By zooming in at small values of the strain, we observe a universal behavior:
\begin{equation}
    |\delta T_c|\,\sim\,\epsilon^2 \,,
\end{equation}
where $\delta T_c$ is the difference between the critical temperature at zero strain and that at finite strain. Such behavior is plotted, for some values of $\rho L_\alpha^2$, in the inset of the left panel of Fig.\ref{fig:3}.

Once more, our numerical results are consistent with the prediction of the effective field theory in Section \ref{effectivesection}, specifically with Eq.\eqref{eq:shear_tc_corr}. Using the values of the coefficient $a_2^{(\mathcal{G})}$ previously extracted, we are now able to obtain the second parameter appearing in the EFT description: $a_0$. We show the value of this parameter in function of the dimensionless charge density in the right panel of Fig.\ref{fig:3}. In the regime considered, we observe a monotonically decreasing behaviour. Unfortunately, our numerical scheme does not allow us to have control over the computations at small $\rho L_\alpha^2$ and we cannot therefore exclude a possible non-monotonic behaviour for small values of the charge density. However, $a_0$ is a parameter of the pure superfluid phase, unlike $a_2^{(\mathcal{G})}$, which describes its interplay with the breaking of translations; therefore, a qualitatively different behavior would not be surprising.\medskip

\begin{figure}
    \centering
    \includegraphics[width=0.88 \linewidth]{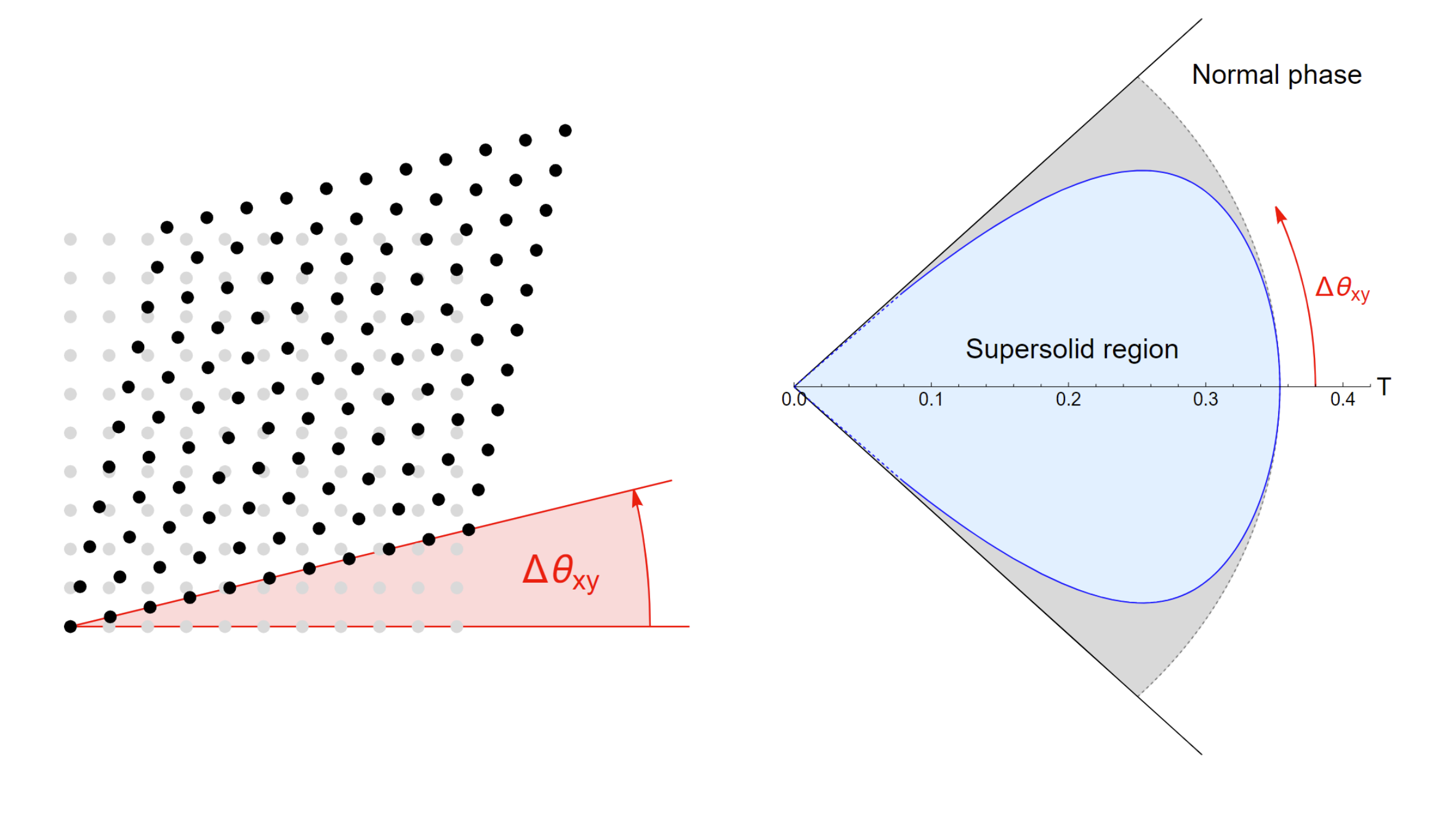}
    \caption{\textbf{Left:} Effects of a pure shear transformation ($\epsilon=0.5$) on a square lattice, and definition of the pure shear angle $\Delta \Theta_{xy}$. \textbf{Right:} An alternative visualization of the phase diagram at fixed charge density ($\rho L_\alpha^2 = 4$) in terms of the shear angle $\Delta \Theta_{xy}$. The blue dashed lines indicates the range of values in which our numerical method loses precision and numerical extrapolation must be used.}
    \label{fig:add3}
\end{figure}

Notice that the curves of the left panel of Fig.\ref{fig:3} provide the same information of the phase diagram of the system at fixed charge density $\rho L_\alpha^2$: indeed every $(T,\, \epsilon)$ point below the appropriate critical curve lies within the supersolid phase, while the ones above in the normal solid phase. Here, we present a more direct way to visualize the finite shear strain by introducing, instead of the shear strain parameter $\epsilon$, the shear angle $\Delta \Theta_{xy}$, whose meaning is illustrated in the left panel of Fig.\ref{fig:add3}. In terms of our original parameter $\epsilon$, which is defined as the off-diagonal component of the deformation gradient tensor $\Xi$, the finite angle can be expressed as:
\begin{equation}\label{eq:shear_angle}
    \Delta \Theta_{xy} = \arctan\left(\frac{\epsilon}{\sqrt{4 + \epsilon^2}}\right)\,.
\end{equation}
Notice that the support of this function is the interval $(-\frac{\pi}{4},\,\frac{\pi}{4})$. This is indicative of the fact that for a pure shear , as opposed to a simple shear, the maximum shear angle, defined as the one where the systems collapses on a line, is $\pm \frac{\pi}{4}$.

Exploiting this relation, we can provide an alternative visualization for the fixed-density phase diagram, presented in the right panel of Fig.\ref{fig:add3} for one $\rho L_\alpha^2$ value. Inside the lobe-shaped curve, the system is supersolid, outside it is in its normal phase. Qualitatively similar curves can be found for different values of the dimensionless charge density.

\medskip

Then, we move to analyze the opposite scenario of a purely volumetric deformation in which the shear strain $\epsilon$ is taken to be zero. A finite volume change $\delta =\Delta V/V$ is parameterized by the bulk parameter $\alpha$ as shown in Eq.\eqref{law}.

As a consequence of keeping constant the combination $\rho L_\alpha^2$, we find that:
\begin{equation}\label{eq:tc_prob}
    T_c \,L_{\alpha_i} = T_c' \,L_{\alpha_f}\quad \longrightarrow \quad \frac{T_c'}{T_c}\,=\,\frac{\alpha_i}{\alpha_f}
\end{equation}
where $T_c'$ is the critical temperature after the deformation, and $T_c$ the one before.

\begin{figure}
    \centering
    \includegraphics[trim=25 0 0 0,clip,width=0.45 \linewidth]{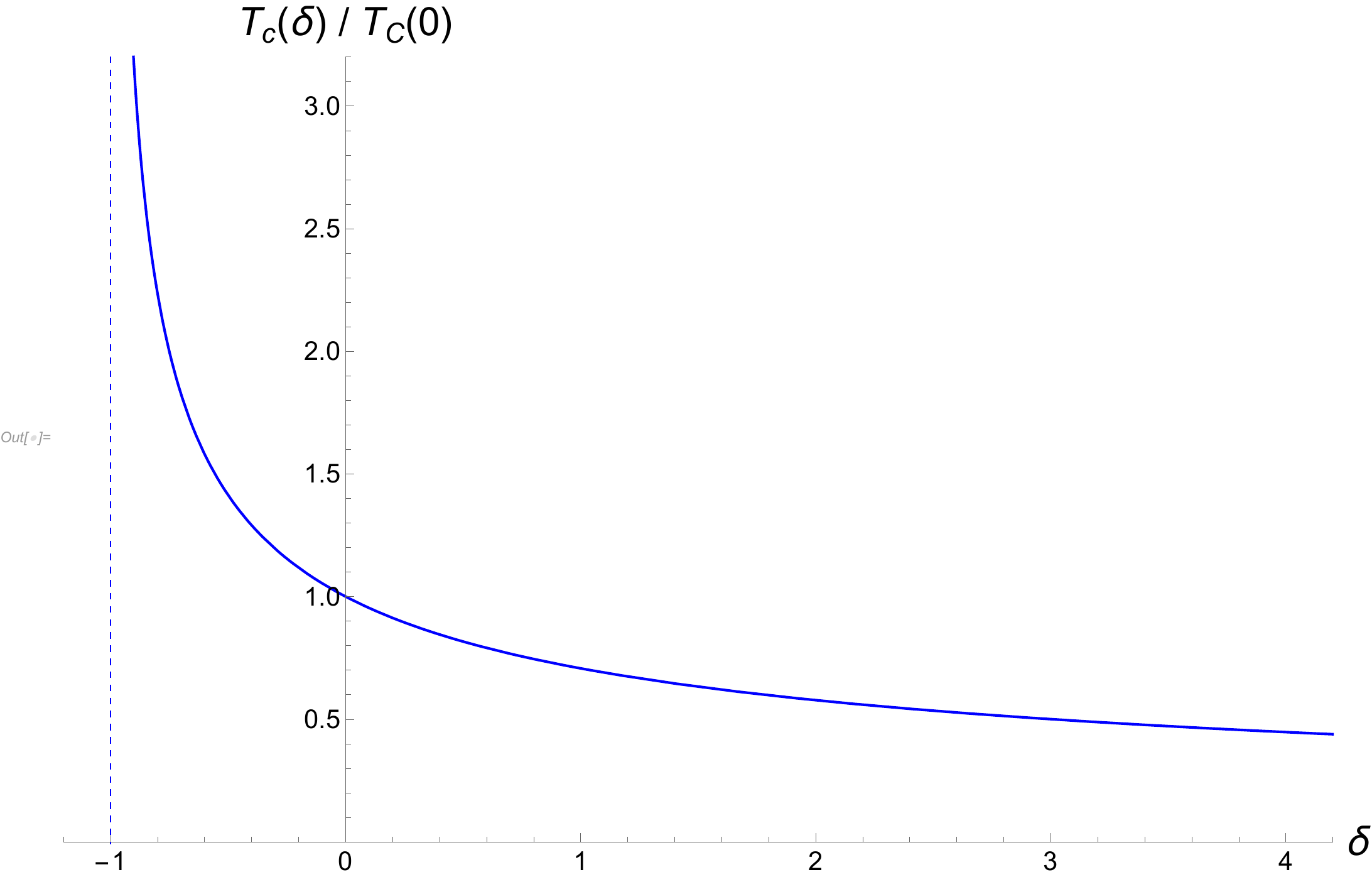} \qquad
    \includegraphics[trim=25 0 0 0 ,clip,width=0.45 \linewidth]{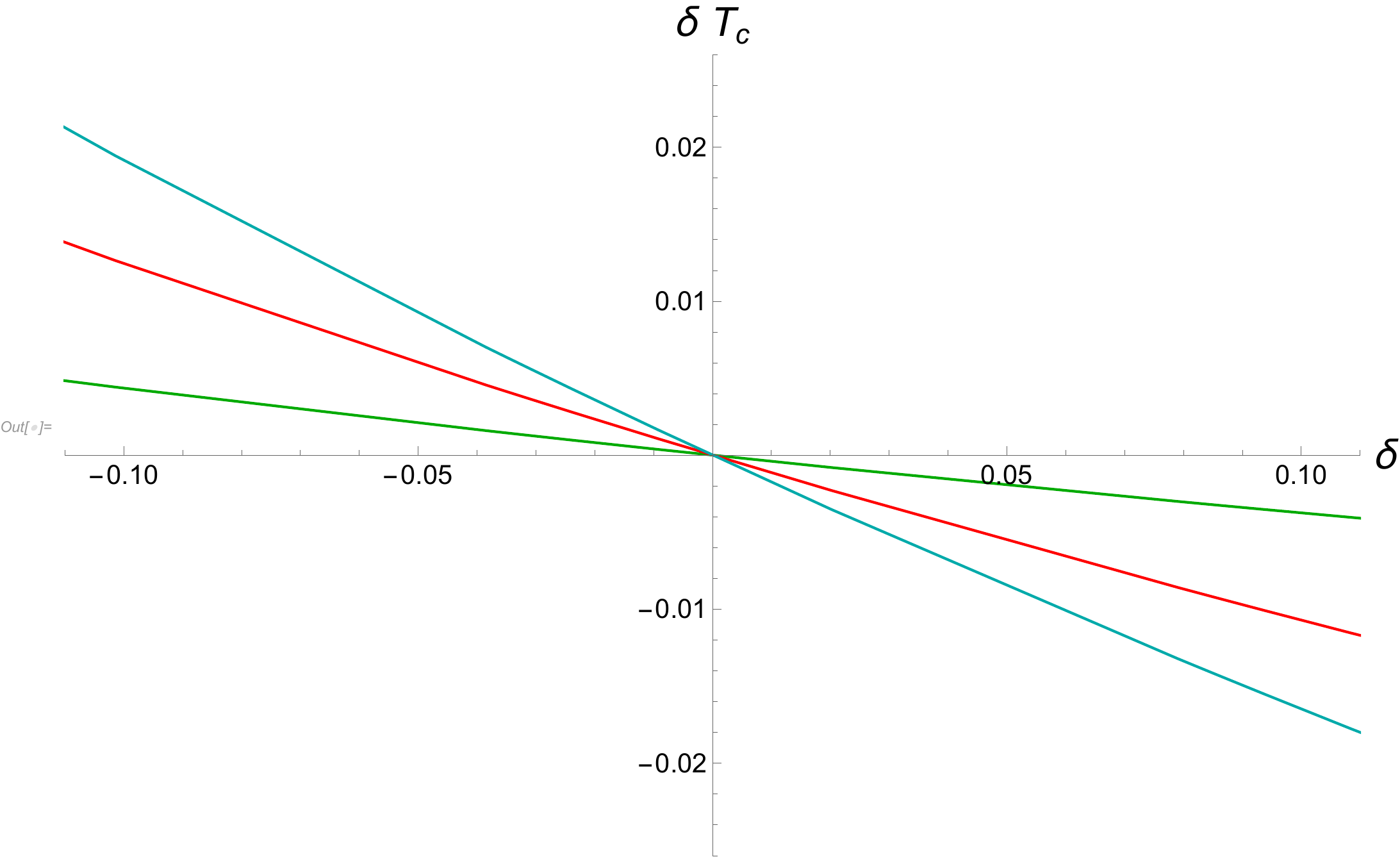}
    \caption{\textbf{Left: } The universal function regulating the behavior of the critical temperature under isotropic compressions ($\delta<0$) and dilatations ($\delta>0$). \textbf{Right: } The corrections to the critical temperature from a purely volumetric deformation $\delta$, for the notable values of adimensional charge density defined in Fig.\ref{fig:1}.}
    \label{fig:4}
\end{figure}

For simplicity, and without loss of generality, let us then consider \eqref{eq:tc_prob} with the reference state $\alpha_i = 1$. One would obtain:
\begin{equation}
    T_c' = T_c \,  \alpha,
\end{equation}
after the deformation, where we have identified $\alpha_f=\alpha$. To compare with the results of Section \ref{effectivesection}, we need to use the volumetric deformation parameter $\delta$ as defined in \eqref{dd1}, obtaining thus:
\begin{equation}\label{eq:bulk_tc_law}
    T_c' = \frac{T_c}{\sqrt{1 + \delta}} \approx T_c \left(1 - \frac{1}{2} \delta + \frac{3}{8} \delta^2 + ... \right),
\end{equation}
where the first half of the equality is valid all orders in $\delta$, while the second half follows from the Taylor expansion of the right hand side for small deformations, $\delta \rightarrow 0$. The fully non-linear results can be then expressed as:
\begin{equation}\label{eq:universal_correction}
    \frac{T_c(\delta)}{T_c(0)} = \frac{1}{\sqrt{1+\delta}}\,,
\end{equation}
plotted in the left panel of Fig.\ref{fig:4}. The behavior of the function suggests that, whenever the system is compressed ($\delta<0$), the critical temperature rises, viceversa, whenever it undergoes a dilatation ($\delta>0$) it decreases. Some notable additional features are the singularity at $\delta=-1$, which is the limit of infinite compression at which the system shrinks to a point, and the fact that the function has no zero; this latter fact provides a good consistency check, as in our model it is the value of $\rho L_\alpha^2$ which decides whether the system must have a supersolid phase or not, regardless of the value of $L_\alpha$; as such, the zero temperature broken phase must resist for all values of the deformation parameter. In the right panel of Fig.\ref{fig:4}, we show the effect of small deformations around the reference configuration for various values of $\rho L_\alpha^2$. The linear slope increases with the charge density $\rho L_\alpha^2$.\medskip

\begin{figure}
    \centering
    \includegraphics[trim=25 0 0 0,clip,width=0.45 \linewidth]{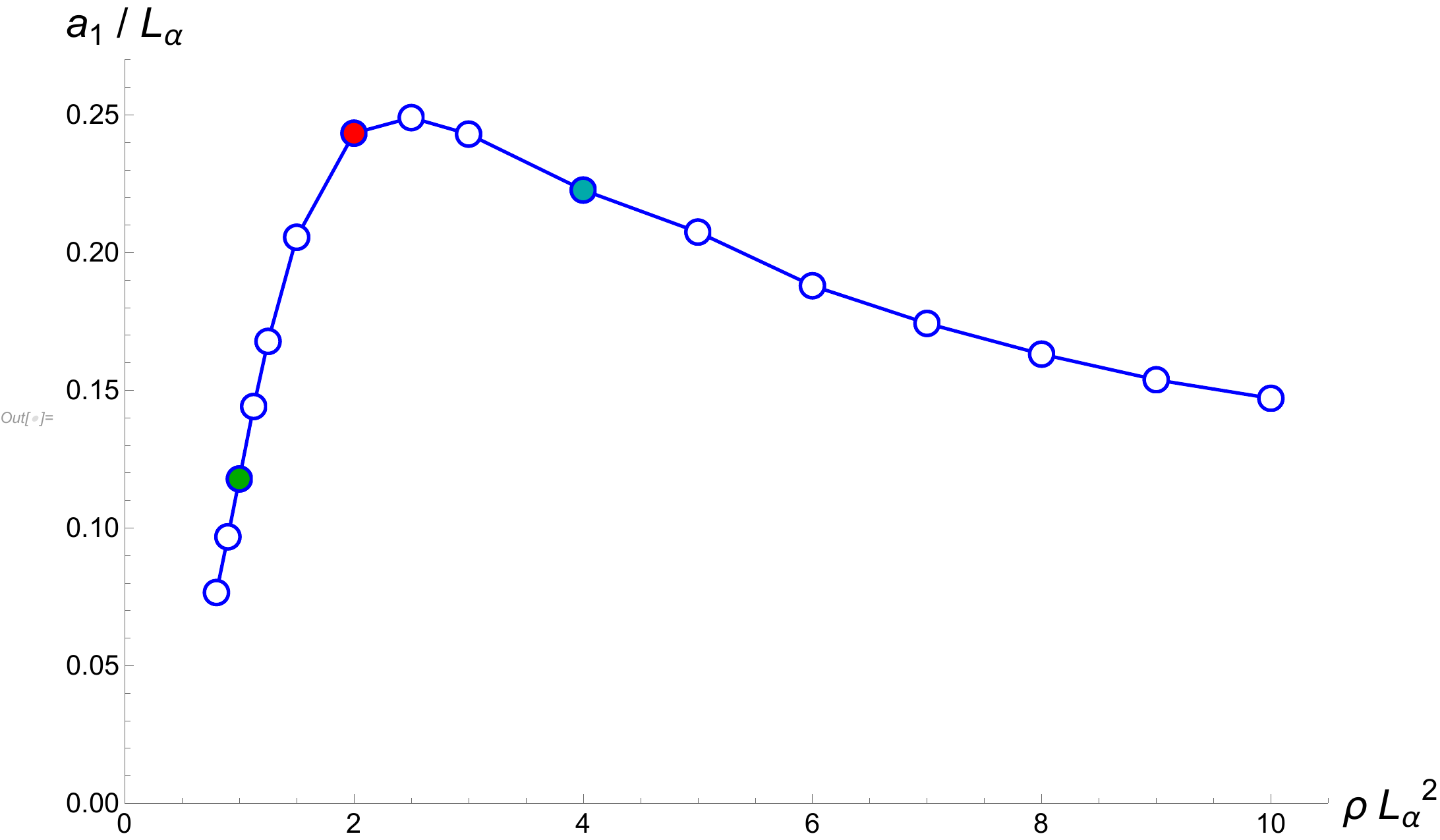} \qquad
    \includegraphics[trim=25 0 0 0 ,clip,width=0.45 \linewidth]{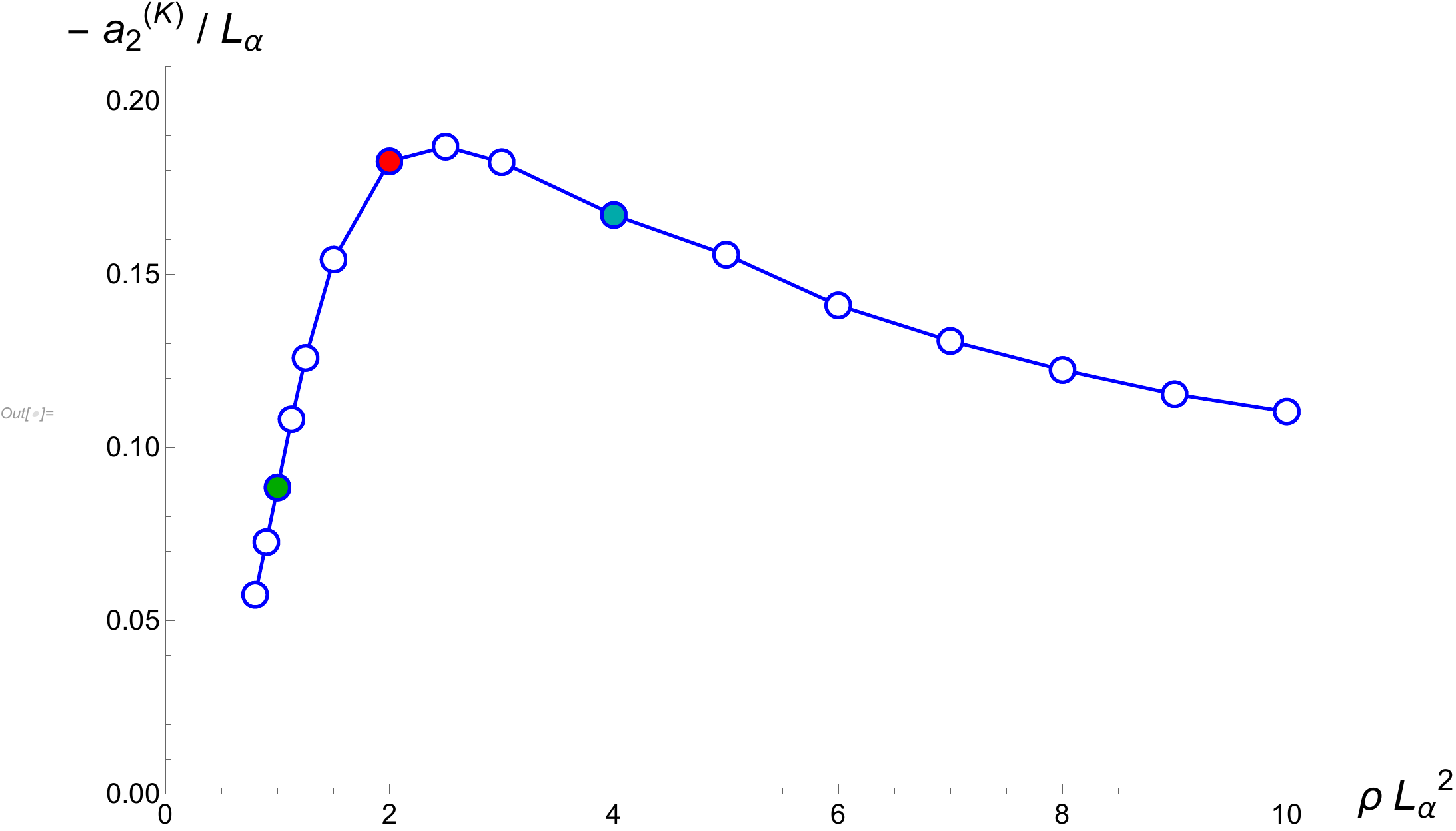}
    \caption{The corrections to the critical temperature from a volumetric strain appearing in Eq.\eqref{eq:delta_tc_corr} in function of $\rho L_\alpha^2$. \textbf{Left: } The linear coefficient $a_1$. \textbf{Right: } The quadratic term $a_2^{(\mathcal{K})}$.}
    \label{fig:5}
\end{figure}

Using the small deformation expansion in Eq. \eqref{eq:bulk_tc_law}, it is possible to extract the coefficients $a_1$ and $a_2^{(\mathcal{K})}$ of the EFT of Section \ref{effectivesection} by comparing it with \eqref{eq:delta_tc_corr}. They are:
\begin{equation}
    a_1 = \frac{1}{2} \, a_0 \, T_c, \qquad a_2^{(\mathcal{K})} =  -\frac{3}{8} \, a_0 \, T_c\,.
\end{equation}
The results are reported in Fig.\ref{fig:5}. The behavior of these coefficients, which is qualitatively the same as one is a multiple of the other, is similar to the one of the $a_2^{(\mathcal{G})}$ coefficient, reported in the right panel of Fig.\ref{fig:2}. The only notable differences come from the fact that the maximum seems to be shifted towards higher values of $\rho L_\alpha^2$, but still around order $\mathcal{O}(1)$, and that it does not appear evident whether or not the coefficient is going to vanish for large values of the adimensional charge density.

\subsection{Interplay between large shear deformations and condensation}

In the previous sections we outlined extensive evidence of the fact that the model at hand exhibits a rich interplay between elasticity and superfluidity. We characterised how deformations (small and large) change the critical properties of the system, and how the condensation -- even far from the transition -- does have an impact on the response to linear mechanical deformations.
In this section, we investigate the fully nonlinear regime, to estimate how large (shear) deformations affect the condensation, and how the stress response to large deformations is changed by condensation.

\subsubsection{Strain-induced transition}

By looking at the curves of Fig.\ref{fig:3}, it seems quite clear that this model predicts the possibility to have an isothermal, shear-induced transition back to the normal state.

\begin{figure}
    \centering
    \includegraphics[trim=25 0 0 0,clip,width=0.7\linewidth]{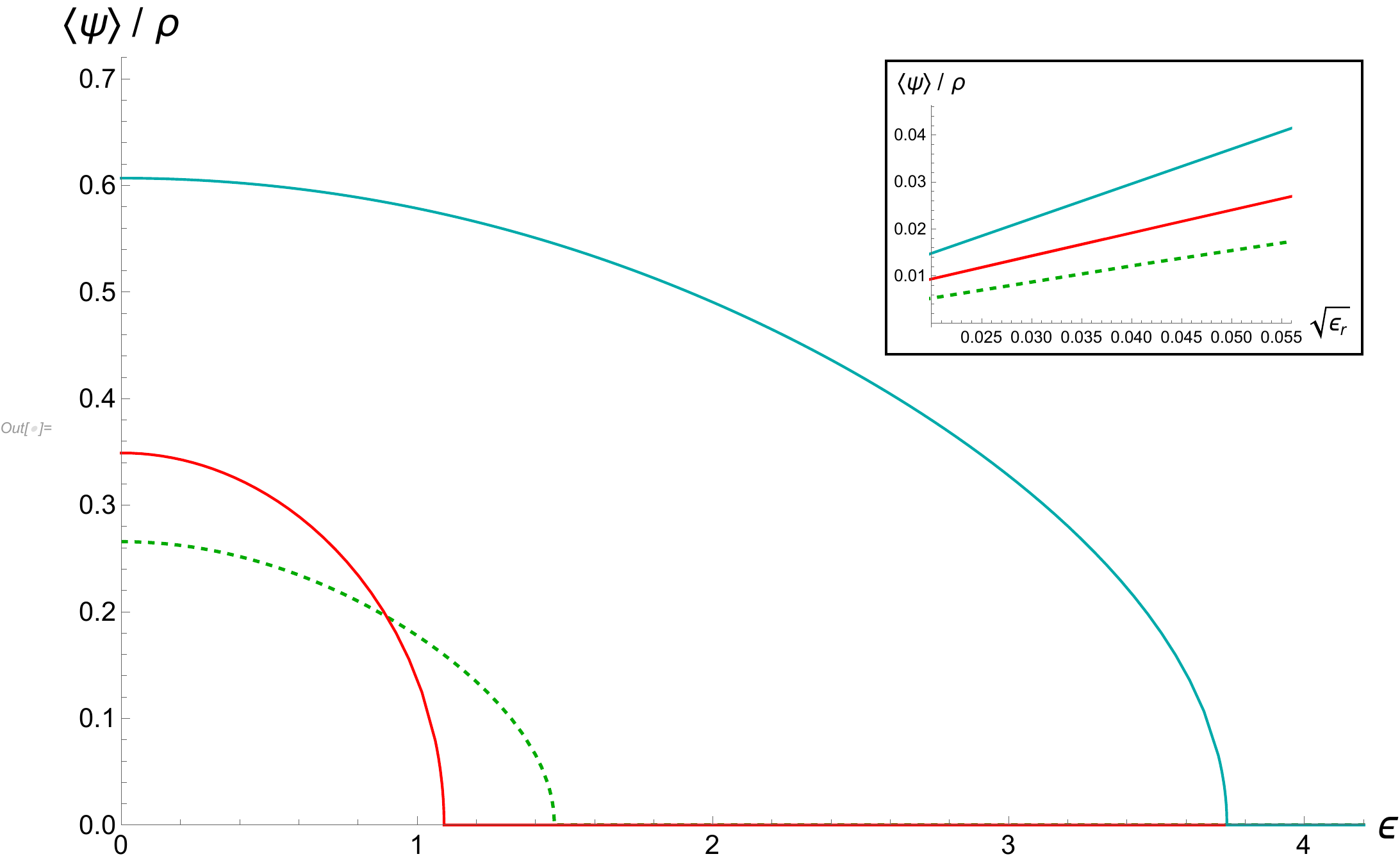}
    \caption{The superfluid condensate throughout an isothermal pure shear deformation process. Solid lines indicate processes at $T L_\alpha = 0.2$, while the dashed line at $T L_\alpha = 0.05$. The endpoints of the curve are in accordance with the ones predicted in Fig.\ref{fig:3}. \textbf{Inset: } The dimensionless condensate close to the strain induced transition, as a function of the square root of the reduced strain parameter $\epsilon_r \equiv |\epsilon-\epsilon_c|/\epsilon_c$ with $\epsilon_c$ the point where superconductivity disappears. The linear behavior of the curve points out the mean-field theory nature of the transition. Notice that in this panel, $\rho$ has been used to adimensionalise the condensate, for the sake of a better graphical representation.}
    \label{fig:6}
\end{figure}

We report the results for such a study in Fig.\ref{fig:6}. We find that, starting from an adimensional temperature below the critical one ($T^* L_\alpha < T_c L_\alpha$) and keeping it fixed throughout the process, the adimensional condensate starts decreasing as the shear strain parameter $\epsilon$ is increased, until a critical value of the latter occurs where the supersolid phase disappears. Such critical value should be identified with the solution of the equation:
\begin{equation}
    T_c L_\alpha \left( \rho L_\alpha, \, \epsilon \right) = T^* L_\alpha,
\end{equation}
and indeed this is what we find, up to the precision of our numerical methods.

Interestingly, we can also study the condensate near the transition, to see what kind of transition the shear-strain induced one is. In the inset of Fig.\ref{fig:6}, evidence that it is a mean field transition is reported. The condensate indeed obeys a square-root law in the reduced shear strain parameter, which we have defined in analogy with the reduced temperature as:
\begin{equation}
    \epsilon_r = \frac{|\epsilon - \epsilon_c|}{\epsilon_c},
\end{equation}
with $\epsilon_c$ being the \textit{critical shear strain parameter} above which no supersolid phase is present.

\subsubsection{Condensate-enhanced stress response}

Another natural study is to extend the results of Section \ref{subsub:shear_mod} to the non-linear regime and to see how the condensation affects the stress response when large deformations are involved. Instead of working with the shear modulus $\mathcal{G}$, which is a linear quantity, we work with the full stress-strain relation. In order to do so, we will generalise the results of \cite{baggioli2020black}, which laid the groundwork for the calculation of such relations in a fully nonlinear, holographic setup.

Let us review briefly what we mean by this. In a linear theory of elasticity -- i.e. when deformations are small --, the effective free energy regulating the processes described by the theory is \eqref{solid}, so that the stress response $\sigma_{\alpha\beta}$ that follows is:
\begin{equation}
    \sigma_{\alpha\beta} = C_{\alpha\beta\gamma\delta}\, \epsilon_{\gamma\delta},
\end{equation}
with $\epsilon_{\gamma\delta}$ the usual linear strain tensor. In the case of shear deformations, for a homogeneous and isotropic medium, and adopting our parameterization, one finds:
\begin{equation}
    \sigma = \mathcal{G} \epsilon,
\end{equation}
where $\sigma = \sigma_{xy}$. When deformations become large enough, the response is described by a more general relation:
\begin{equation}
    \sigma = \sigma(\epsilon),
\end{equation}
where now $\sigma(\epsilon)$ is some generic nonlinear function of the parameter $\epsilon$, that we expect to be linear in the limit $\epsilon \rightarrow 0$. Holographically, $\sigma$ can be shown to be proportional to the subleading term of the UV expansion of the $h(u)$ function appearing in the metric. Such expansion can be found to be, in our case of interest:
\begin{equation}
    h(u) \simeq h_3 \, u^3 \, (1 + ...).
\end{equation}
\begin{figure}
    \centering
    \includegraphics[trim=25 0 0 0,clip,width=0.7\linewidth]{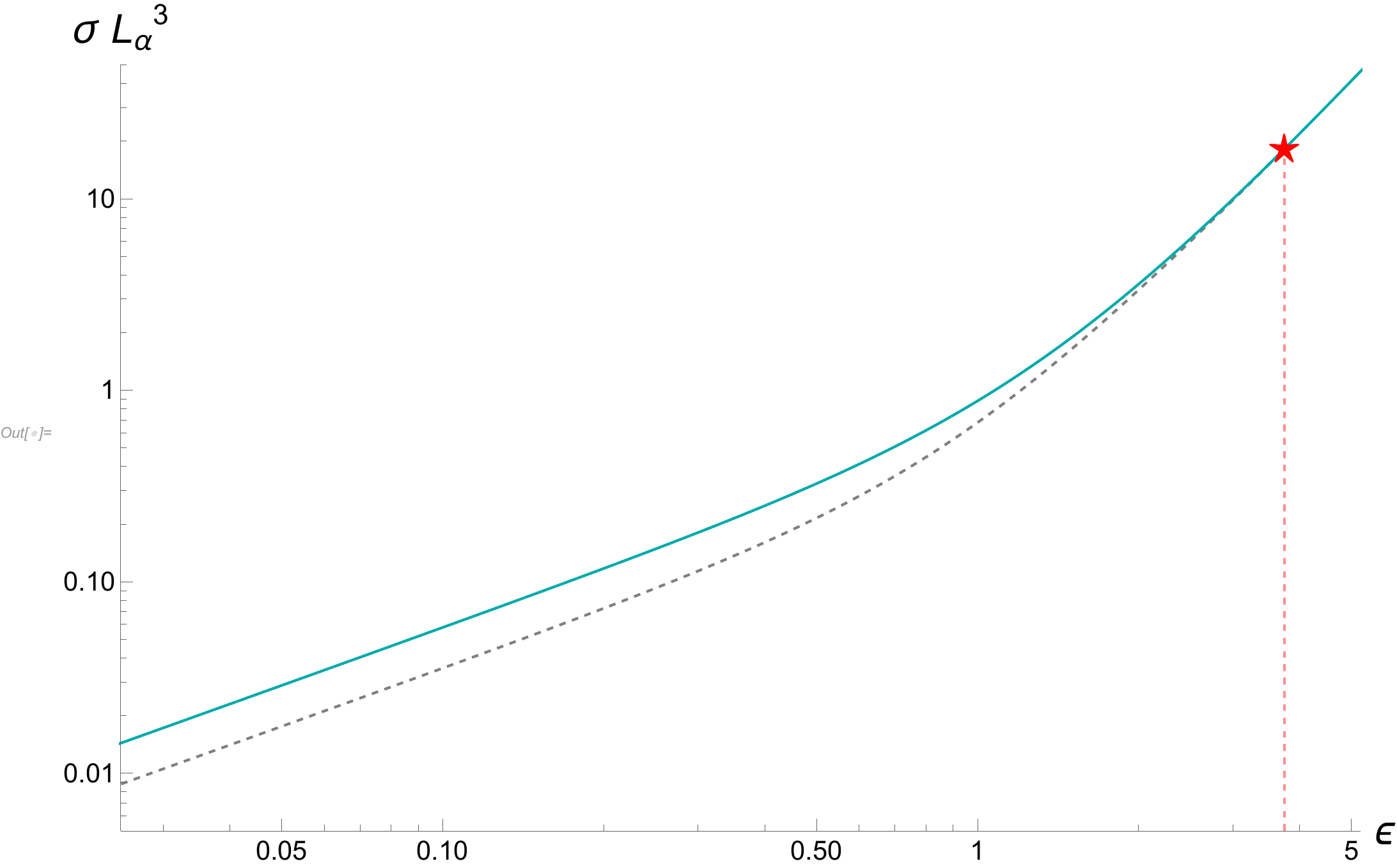}
    \caption{The stress-strain curve for $T L_\alpha = 0.2$ and $\rho L_\alpha^2 = 4$. The dashed line represent the would-be mechanical response in absence of condensation (i.e. in the solid phase without superfluid component). The red star indicates where the two curves join -- at the critical strain parameter $\epsilon_c$, whose value is further stressed by the vertical red, dashed line.}
    \label{fig:stress-strain}
\end{figure}
We refer to the original work \cite{baggioli2020black} for further details. Here, we limit ourselves to report the useful final formula:
\begin{equation}
    \sigma = \frac{3}{2} \, h_3,
\end{equation}
that can be used to compute the shear stress numerically. Fig.\ref{fig:stress-strain} reports one comparison between the stress-strain relation in the supersolid and solid states. The results are consistent with what found before: the presence of the condensate enhances the stress response. Clearly, this enhancement vanishes as the system transitions back into the normal phase due to the increased strain.

\section{Outlook}
\label{sec:conclusions}
In this work we have initiated the study of supersolid phases of matter using the holographic framework by considering solutions which break spontaneously translations together with a global $U(1)$ symmetry. 

The first interesting observed feature is an increase of the shear modulus in the supersolid phase which can be rationalized as a contribution from the superfluid component. As expected from the effective theory description \cite{PhysRevLett.96.055301} reviewed in Section \ref{effectivesection}, close to the supersolid critical temperature the increase is linearly proportional to the square of the superfluid condensate  $|\Psi|^2$. This result is compatible with the experimental results appeared so far in the literature \cite{Day2007,doi:10.1063/1.2883895,Syshchenko_2009}. The same behavior -- with opposite sign though -- has been found by studying the shear viscosity. The latter further decreases in the supersolid phase and presents a discontinuous derivative at the supersolid critical point. Qualitatively, this behavior is in accordance with the notion that, upon a superfluid transition, the normal component of the system, and therefore the associate friction, decrease. Additionally, we find the Kovtun-Son-Starinets bound violation is enhanced by the presence of a condensate and we indicate as a potential responsible for this fact the interactions between the solid and superfluid degrees of freedom. Interestingly, in absence of the solid component and these new interactions, no effects on the $\eta/s$ ratio was found in the literature (at least for the $s$-wave case) \cite{Natsuume_shearvisc}.

Moreover, we have characterized the nature of the supersolid phase transition as a function of the mechanical deformation parameters. In the linear elasticity regime -- for small deformations -- the qualitative behavior of the critical temperature is once more the one predicted by the effective field theory (Section \ref{effectivesection}). We find that a shear strain renders the critical temperature smaller and therefore it tends to destroy the supersolid phase. On the contrary, we observe that the behavior in terms of a volumetric deformation depends on its sign. A volumetric compression increases the supersolid critical temperature, while an isotropic expansion disfavours the broken phase.

Finally, we have studied the regime of large mechanical deformations and how they do affect the supersolid critical temperature. We observe that the superfluid condensate goes to zero at a critical value of the strain following a mean-field behaviour. On the other hand, the condensed system appears to be stiffer even when deformation become large.\medskip

An obvious extension of our work would be to build a relativistic hydrodynamics framework for supersolids and compare it with our holographic models. For example, it is well known that a supersolid supports two different longitudinal sound modes with different speeds of propagation \cite{PhysRevLett.108.175301}. All this information can be obtained by extracting numerically the quasinormal modes of our system. It would be also interesting to think about apply an external rotation to our supersolid system in order to check in more detail how close is the holographic model to what expected for real supersolid systems.\medskip

Let us conclude with two comments about the phenomenology of our holographic toy model. Pair-density waves (PDWs) have been constructed holographically in \cite{Cai:2017qdz,Cremonini:2017usb,Cremonini:2016rbd}. To the best of our knowledge, without appealing to microscopic details, the symmetry properties of PDWs and supersolids are very similar -- they both break spatial translations together with a global $U(1)$ symmetry. Therefore, at least at low energy (late time and large scales) we do not expect major differences between our findings and the physics of PDWs. It would be fruitful to make this point clearer and discuss the commonalities/differences between our model and results and the holographic setups for PDWs mentioned above.
Another object of prospective further studies is the question regarding strain-induced instabilities. In real-world materials, stress-strain curves are known to have an endpoint -- usually corresponding to the point where the sample breaks --.
A first study on how to determine these \textit{elasticity bounds} in zero temperature effective field theories is included in the analysis of \cite{baggioli2020black} and used afterwards in \cite{Pan:2021cux}. Mechanisms proposed there as possible signatures of strain-induced instabilities are imaginary speeds of sound in the bulk or the appearance of ghosts. Such proposals, though, have been brought forward in the context of the \textit{decoupling limit}; the only reliable way to obtain such bounds -- at least numerically -- would be to check whether the dispersion relation of perturbations around the large-strained state develop any pathology. This requires a considerable effort and is left for future investigations. From a holographic perspective, the analysis should be analogous to that of holographic superfluids with background superfluid velocity \cite{Amado:2013aea}. Importantly, this approach could put important bounds on supersolidity and/or shed some lights on their instability properties in the nonlinear regime \cite{PhysRevLett.104.075302}.\\

Finally, despite it is well-known (but too often ignored) that the HHH model \cite{Hartnoll:2008vx} does not represent the gravity dual of a superconducting state of matter (unless one imposes mixed boundary conditions for the bulk gauge field \cite{Domenech:2010nf}), we could still ask ourselves whether our results might be relevant to discuss the impact of mechanical deformations in the context of superconductors, and especially non BCS-like ones. Unfortunately, it seems that no universal trend has emerged so far and all possible behaviors have been experimentally observed \cite{buehler1965effect,zhai2000effect,medvedev2009electronic,han2010superconductivity,qing2012interface,hicks2014strong,steppke2017strong,ahadi2019enhancing,ruf2021strain,ghini2021strain,locquet1998doubling,malinowski2020suppression}. By abusing of the model and considering the critical temperature as the superconducting (SC) one, our computations suggest that a compression always increases the SC critical temperature while a dilatation or a shear strain decrease it. It would be interesting in the future to assess whether the holographic methods could provide any useful information in this direction as well and compare to the experimental trends reported above.

\section*{Acknowledgments}
We especially thank Alessio Zaccone and Oriol Pujol{\`a}s for their collaboration at an early stage of this project and for uncountable suggestions and discussions.
We thank Wei-Jia Li and Li Li for several discussions on the topics of this work and for providing useful comments on a first version of this manuscript. We thank Blaise Gout\'eraux for a related collaboration which helped a lot in clarifying several aspects of this work. We thank Sašo Grozdanov and the Holography, Strings and Transport group at University of Ljubljana for the logistic support in these hard times. M.B. acknowledges the support of the Shanghai Municipal Science and Technology Major Project (Grant No.2019SHZDZX01). G.F. is profoundly thankful to his parents for their loving support, and to all the academics that provided him with their precious mentorship throughout the time, especially to those in Milan.

\appendix

\section{A brief review of finite elasticity theory} \label{app1}
In this Appendix, we briefly review some basics of finite elasticity theory following the arguments and notations of \cite{ZAMM:ZAMM19850650903}. For the sake of the present work, we will restrict our attention to non-dissipative processes and materials with no defects. In such framework, one can see any deformation process as a smooth, invertible map from one initial (\textit{reference}) to a final (\textit{current}) configuration. \medskip

Given a reference frame, one way to uniquely identify any geometrical configuration is by specifying the position of every of its constituents -- be they atoms or more generic infinitesimal volume elements --. This can be obtained by introducing a set of scalar fields $\varphi^I(x^i)$, often referred to as \textit{crystal fields} \cite{Armas:2019sbe}, in a number equal to the spatial dimensions of the system\footnote{Strictly speaking, the crystal fields could be \textit{less} than the number of dimensions, as it is the case for smectic crystals; for isotropic solids, though, the equality must hold.}. Symbolically:
\begin{equation*}
    \text{Field profile } \varphi^I \Longleftrightarrow \text{Material configuration}\,.
\end{equation*}
Notice that we used a different convention for the index of the coordinates $x^i$ (lowercase Latin) and of the crystal fields $\varphi^I$ (uppercase Latin). The meaning of such distinction will be clarified in short.

With this in mind, we proceed to give a more precise mathematical formulation of our problem. As we said, a deformation is a regular map going from one configuration to the other. Because we are free to pick our reference frame, a very natural choice is to choose one where the reference configuration \textit{coincides} with the coordinates, in the sense that:
\begin{equation}\label{eq:reference_state}
    \varphi_0^I(x^i) = {\delta^I}_i\,x^i.
\end{equation}
where $x_j$ are the (spatial) coordinates adopted. Notice that in doing so we did not state anything regarding the nature of this reference configuration; rather, we limited ourselves to pick a convenient framework to work with. In most applications, such a configuration is assumed to be the equilibrium one (for which the total stress vanishes), but there is no necessity to assume this, so that the formalism presented here is valid even outside of such hypothesis.

Another important observation regarding \eqref{eq:reference_state} is that in such a state there exist virtually no distinction between upper- and lowercase indices. This is a feature of the choice of reference frame, and we learn that, until we stick with this convention, lowercase indices are associated to the reference configuration. In the literature, they are also called \textit{Lagrangian}. In more general terms, the \textit{Lagrangian framework} of elasticity is the one where the current configuration is expressed as a function of the reference one, while the opposite standpoint -- made possible by the invertibility of each deformation map -- is referred to as \textit{Eulerian}. \medskip

\begin{figure}[t]
    \centering
    \includegraphics[trim=25 0 0 0,clip,width=\linewidth]{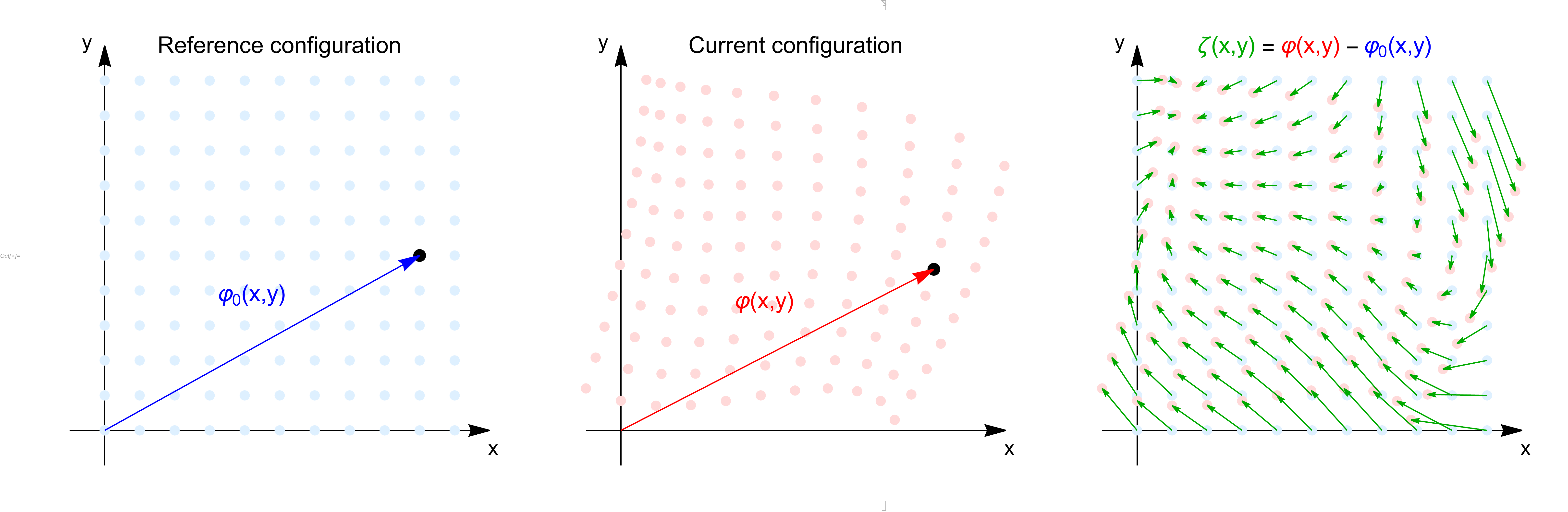}
    \caption{A representation of the reference configuration $\varphi_0^I=x^I$ (left), the current configuration $\varphi^I$ (center) and the displacement field $\zeta_I$ (right).}
    \label{fig:scheme}
\end{figure}

In elasticity theory, the fundamental object is the \textit{displacement vector}, a vectorial field that specifies where each of the points is mapped when a deformation occurs. In our language, this is equivalent to subtracting the reference configuration from the current one $\varphi^I(x^i)$. In flat space, this gives the displacement vector:
\begin{equation}
    \zeta^I (x^i) \equiv \varphi^I(x^i)-{\delta^I}_j \, x^j\,,
\end{equation}
where we assumed a Lagrangian point of view. A representation of this framework is provided in Fig. \ref{fig:scheme}. \medskip

In the theory of finite deformations, a prominent role is played by the mixed \textit{deformation gradient tensor}, defined as:
\begin{equation}\label{e2}
    {\Xi^I}_j \equiv \frac{\partial \varphi^I}{\partial x^j}.
\end{equation}
This is nothing but the Jacobian of the transformation, which can be used as a building block to construct various (nonlinear) versions of the strain tensor. Two of those can be obtained by comparing the distance between two neighboring points before and after the deformation. We have, indeed, that:
\begin{equation}\label{e3}
    d \varphi^I = {\Xi^I}_j \, d x^j,
\end{equation}
together with an inverse relation. Continuing on these lines, we can define the \textit{Green-Lagrange strain tensor} $E_{ij}$ via the relation:
\begin{equation}\label{e1}
    d\varphi^2-dx^2\,\equiv  dx^T\,\cdot\,2\,E\,\cdot dx.
\end{equation}
Using Eq.\eqref{e1} together with Eq.\eqref{e2}, we then obtain:
\begin{equation}
    E\,=\,\frac{1}{2}\left(\Xi^T \Xi-\mathbb{I}\right)\,,\label{efull}
\end{equation}
where $\mathbb{I}$ is the identity. In components, we have:
\begin{equation}
    {E^i}_j=\frac{1}{2}\left({\Xi^i}_{K}{\Xi^K}_j - {\delta^i}_j \right) ,
\end{equation}
so that we are left with Lagrangian indices only. In terms of the displacements -- with an abuse of notation, all the indices have become Lagrangian -- the relation reads:
\begin{equation}
    E_{ij}=\frac{1}{2}\left(\partial_i \zeta_j +\partial_j \zeta_i+ \partial_i \zeta_k\, \partial_j \zeta_k\right).\label{thise}
\end{equation}
An analogous procedure can be followed to obtain a Eulerian strain tensor. It will suffice to use the inverse of \eqref{e3} in the defining relation:
\begin{equation}
    d\varphi^2-dx^2\,\equiv  d\varphi^T\,\cdot\,2\,E'\,\cdot d\varphi.
\end{equation}
One obtains:
\begin{equation}
    E'\,=\,\frac{1}{2}\left(\mathbb{I} - (\Xi^{-1})^T \Xi^{-1} \right)\,, \label{efulleuler}
\end{equation}
which has indeed two Eulerian indices. Its expression in terms of the (Eulerian) displacement vector $\Bar{\zeta}$ is:
\begin{equation}
    E_{IJ}=\frac{1}{2}\left(\partial_I \Bar{\zeta}_J +\partial_J \Bar{\zeta}_I - \partial_I \Bar{\zeta}_K\, \partial_J \Bar{\zeta}_K\right).
\end{equation}

It is worth noticing that at the linear level, when quadratic terms in the displacements are negligible, both \eqref{efull} and \eqref{efulleuler} reduce to the linear strain tensor $ \epsilon_{ij}$ typically used in linear elasticity theory \cite{chaikin_lubensky_1995}. This can be seen as a consequence of the fact that, in that regime, the difference between the two standpoint blurs, as the transformations themselves are infinitesimal. In any case, the key lesson to be learned here is that in the nonlinear regime multiple definitions for the strain tensor are available, and which to use reduces to a matter of mathematical convenience. In this work, we adopt the Green-Lagrange one \eqref{efull}. \medskip

To conclude this appendix, let us analyze in more detail the structure of the deformation gradient tensor $\Xi$ in light of some consistency and symmetry-imposed restrictions. For convenience, let us stick to two spatial dimensions, where we would expect, on general grounds, such an object to carry $4$ independent degrees of freedom. Importantly, there is a redundancy in this description, which reduces their number. Let us see why. Because $\Xi$ is a non-singular matrix, the polar decomposition theorem holds, which means that there exist two (uniquely defined) symmetric matrices $U$ and $W$, and a rotation matrix $R$ such that:
\begin{equation}\label{eq:polar_decomposition_theorem}
    \Xi = U \cdot R = R \cdot W.
\end{equation}
Using the definition of the strain tensor in Eq.\eqref{efull}, and the property that $ R R^T=I$, it is straightforward to verify that:
\begin{equation}\label{eq:energy_polar_decomposition}
    E(\Xi)\,=\,E(U)\,=\,E(W)\,.
\end{equation}
This implies that all the matrices $\Xi$, $U$ and $W$ give the same strain tensor: from the point of view of mechanical deformations, they are completely equivalent\footnote{We remind that the energy functional depends exclusively on the strain tensor and not on the deformation gradient tensor. Another way to see this is by recurring to the notion of \textit{material objectivity}, mentioned in \cite{ZAMM:ZAMM19850650903}.}. In operative terms, the equivalence relation defined in Eq.\eqref{eq:energy_polar_decomposition} allows us to always choose a symmetric deformation gradient tensor with only $3$ independent parameters. \medskip

In this work, we deal with isotropic systems. This notion allows us to push this reduction further using the invariance of the systems under rotations by defining an equivalence class under the transformation:
\begin{equation}
    \Xi' = R^{T} \cdot \Xi \cdot R,
\end{equation}
which tells us that all deformation gradient tensors are equivalent up to a rigid rotation $R$. Using this assumption, the resulting number of independent parameters appearing in the gradient deformation tensor is reduced to $2$ (in two dimensions). \medskip

In case of homogeneous deformations, then, the deformation gradient $\Xi$ is constant. Convenient parameterizations for matrices with the same symmetry properties that used in this manuscript are:
\begin{equation}
   \Xi = \alpha
     \begin{pmatrix}
        \cosh{(\Omega / 2)} & \sinh{(\Omega / 2)} \\
        \sinh{(\Omega / 2)} & \cosh{(\Omega / 2)}
    \end{pmatrix} = \alpha 
     \begin{pmatrix}
        \sqrt{1 + \epsilon^2/4} & \epsilon / 2 \\
        \epsilon / 2 & \sqrt{1 + \epsilon^2/4}
    \end{pmatrix} \,,
\end{equation}
where $\alpha$ parameterize a purely volumetric (bulk) deformation in case $\Omega$ vanishes, while $\Omega$ (with $\alpha = 1$) a purely deviatoric term which modifies the shape of the material but not its volume ($\mathrm{det}\left[\Xi\right]=1$). \medskip

From the last equality, it is possible to derive the parametrization of the linear strain tensor shown in the main text in Eq.\eqref{eq:parameterization} by simply expanding \eqref{efull} near the reference configuration $\Xi = \mathbb{I}$. An efficient way to do so is by operating the substitution:
\begin{equation}
    \Xi = \alpha 
     \begin{pmatrix}
        \sqrt{1 + \epsilon^2/4} & \epsilon / 2 \\
        \epsilon / 2 & \sqrt{1 + \epsilon^2/4}
    \end{pmatrix} \longrightarrow (1 + \tilde \delta) 
     \begin{pmatrix}
        \sqrt{1 + \epsilon^2/4} & \epsilon / 2 \\
        \epsilon / 2 & \sqrt{1 + \epsilon^2/4}
    \end{pmatrix},
\end{equation}
and consider both $\tilde\delta$ and $\epsilon$ as small parameters. Then, inserting the new expression into \eqref{efull}, one gets:
\begin{equation}
    E = \frac{1}{2} \left( (1 + \tilde \delta)^2 
     \begin{pmatrix}
        1 + \epsilon^2/2 & \epsilon\sqrt{1 + \epsilon^2/4} \\
        \epsilon\sqrt{1 + \epsilon^2/4} & 1 + \epsilon^2/2
    \end{pmatrix} - \mathbb{I} \right) = \begin{pmatrix}
       \tilde\delta & \epsilon/2 \\
       \epsilon/2 & \tilde\delta
    \end{pmatrix} + O(\tilde\delta^2,\epsilon^2, \tilde\delta\epsilon).
\end{equation}
Keeping the linear order in the deviation parameters gives the linear strain tensor. To arrive to \eqref{eq:parameterization} one just need to recall that:
\begin{equation}
    \delta \equiv \Tr(u) = 2 \tilde\delta.
\end{equation}
Using this relation one gets to the final form \eqref{eq:parameterization} reported in the main body of this work.\\

Importantly, this also shows that the $\epsilon$ parameter appearing in the linear strain tensor coincides with the off-diagonal component of the nonlinear deformation gradient $\Xi_{xy}$ as stated in the main text as well.

\section{Details of the holographic model}\label{app2}
This appendix is devoted to explain some details of the holographic model discussed in the main text. The covariant equations of motion of the system can be retrieved by applying the usual Euler-Lagrange procedure to the action \eqref{eq:full_action} with respect to the fields $\phi^I$, $\psi$, $A_\mu$, and $g_{\mu\nu}$. One obtains:
\begin{gather}
    \nabla \left(\frac{\partial V}{\partial X} \nabla \phi^I \right) = 0, \label{eq_stuck} \\[4pt]
    \nabla^2 \psi - i q \left[ (\nabla \cdot A) + 2 (A \cdot \nabla) \right] \psi - (q^2 A^2 + M^2) \psi = 0,\\[4pt]
    \nabla_\nu F^{\nu\mu} - e_0^2 \left[ i q \left( \psi^* \partial^\mu \psi - \psi \partial^\mu \psi^* \right) + 2 q^2 |\psi|^2 A^\mu \right] = 0, \\[4pt]
    \begin{split} & R_{\mu\nu} - \frac{1}{2} g_{\mu\nu} \left( R - 2 \Lambda \right) - 2 m^2 \left(\frac{\partial V}{\partial g^{\mu\nu}} - \frac{1}{2} g_{\mu\nu} V(X) \right) - \frac{\kappa^2}{e_0^2} \left( F_{\mu\sigma} F_{\nu}^\sigma - \frac{1}{4}g_{\mu\nu} F^2 \right) - \\[1pt] & - \kappa^2 \left[ \left( \partial_\mu \psi - i q A_\mu \psi \right) \left( \partial_\nu \psi^* + i q A_\nu \psi^* \right) + (\mu \leftrightarrow \nu ) - g_{\mu\nu} |(\partial - i q A)\psi|^2 - g_{\mu\nu}M^2 |\psi|^2 \right] = 0,
    \end{split}
\end{gather}
where $\nabla$ is defined to be the covariant derivative associated to the metric connection. As argued in the main text, \eqref{eq_stuck} is solved by \eqref{eq:def_ansatz}, and to restrict to the homogeneous, isotropic case we adopt the ansatz \eqref{eq:rest_ansatz} for the rest of the bulk fields. After writing $\psi(u) = |\psi(u)|e^{i \theta(u)}$, one can readily check that the radial component of Maxwell's equation reads:
\begin{equation}
    \theta'(u) = 0\,,
\end{equation}
so that $\theta(u) = const$ and we may make a phase choice for the condensate field. This reflects the fact that only phase differences are physical, and give rise to measurable phenomena (Josephson effect). Thus, we pick $\theta(u) = 0$ and work with a real scalar field. Following the notations presented in the main text, then, the bulk equations of motion are given by:
\begin{align}
    &  \psi '' + \left(\frac{f'}{f}-\frac{\chi '}{2}-\frac{2}{u}\right) \psi ' + \left(\frac{q^2 A_t^2 e^{\chi }}{f^2}-\frac{L^2 M^2}{u^2 f}\right) \psi =0, \nonumber\\
    &  A_t'' + \frac{\chi'}{2}\, A_t' -\frac{2 e_0^2 L^2 q^2 \psi ^2}{u^2 f} A_t = 0\nonumber \\
    & 2 u f' h'+f \left(2 u h''-h' \left(u \chi '+4\right)\right)+\frac{4 \alpha ^2 m^2 u V_X(X)}{L_0^2}\sinh (\Omega -h) = 0,\nonumber\\
    & -\frac{2 \kappa ^2 q^2 u A_t^2 e^{\chi } \psi ^2}{f^2}-\frac{1}{2} u h'^2-2 \kappa ^2 u \psi '^2+\chi '=0, \nonumber\\
    & -\frac{\kappa ^2 u^4 e^{\chi } A_t'^2}{e_0^2 L^2}+2 u f'-u f \chi '-6 f-2 L^2 m^2 V(X)-2 \kappa ^2 L^2 M^2 \psi ^2+6 = 0,\label{eq:EOMS}
\end{align}
where we used that:
\begin{equation}
    X = \frac{\alpha^2 \, u^2}{L_0^2 L^2} \cosh(\Omega - h).
\end{equation}
The arguments of the functions involved are not reported for readability's sake and primes indicate differentiation with respect to the radial coordinate $u$, while the $X$ subscript with respect to $X$.\medskip

It is vital to notice that the bulk equations of motion \eqref{eq:EOMS} do possess several symmetries, which turn out to be useful for both practical computations and to shed light on some physical ambiguities as well.
\begin{itemize}
    \item The first symmetry is given by the transformation:
\begin{equation}
    e^\chi \rightarrow a^2 e^\chi, \hspace{12pt} t \rightarrow a t, \hspace{12pt} A_t \rightarrow \frac{1}{a} A_t\,.
\end{equation} This is a $\chi$-shift symmetry, which can be used to set $\chi_{UV}\equiv \chi(u=0)=0$ and identify the bulk time coordinate with the boundary one. As argued in the main text, this can be used to reduce the number of physical parameters to be specified to select one particular solution.
\item  A second symmetry is given by:
\begin{equation}
    \begin{gathered}
    \{t,x,y,u\} \rightarrow a \{t,x,y,u\} \\ \Rightarrow \{L,\kappa\} \rightarrow a \{L,\kappa\}, \, \{m,M,A_t,\psi\} \rightarrow \frac{1}{a} \{m,M,A_t,\psi\},    
    \end{gathered}
\end{equation}
which we use to conveniently set the AdS length $L$ to $1$. 
\item The last transformation is:
\begin{equation}\label{eq:scaling_symm_3}
\begin{gathered}
    \{t,x,y,u\} \rightarrow a \{t,x,y,u\} \\ \Rightarrow L_0 \rightarrow a L_0, \, A_t \rightarrow \frac{1}{a} A_t,
\end{gathered}
\end{equation}
which leaves all fields invariant, as well as $X$. From a practical point of view, this can be understood as the possibility of consistently setting $L_0=1$ throughout the calculations; importantly, the way $L_0$ scales in this symmetry is what suggests us that $L_0$ is indeed a length. There is also an additional redundancy in the sense that, at the level of the equations \eqref{eq:EOMS}, the following is a symmetry as well:
\begin{equation}
    L_0 \rightarrow a L_0 \Longleftrightarrow \alpha \rightarrow \frac{1}{a} \alpha\,.
\end{equation}
\end{itemize}

Finally, let us comment on the numerical method used to solve the bulk equations and construct the gravitational solution.

We use a shooting method that starts integrating the EOMs from the black hole's horizon $u_h$ to the UV boundary, where the holographic dictionary allows us to read the physical quantities we are interested into. By examining such equations in the vicinity of the horizon -- and requiring the fields to be regular on it --, we discover that they define a map:
\begin{equation}\label{eq:solution_map1}
    \left(\psi(u_h),\,A_t'(u_h),\, \chi(u_h),\, h(u_h), \,u_h; \,\alpha, \,\Omega,\, L_0 \right) \longrightarrow \text{ Bulk fields},
\end{equation}
where the three parameters on the right of the semicolon are the ones that define which axion field profile we are considering. Once all eight parameters on the left hand side of the map are picked, a unique profile for the bulk fields is selected. By using the above symmetries, it is possible to considerably lower the amount of free parameters to be specified.

To begin, one has to restrict the space of physical solutions by requiring the following conditions:
\begin{enumerate}
    \item $\psi^{(1)}=0$ along with $\psi^{(2)}\neq 0$ (in the broken phase). This corresponds to the condensate assuming a nonzero vacuum expectation value in absence of a source, i.e. to the spontaneous global-$U(1)$ breaking.
    \item $h(0) = 0$, so that the boundary metric is Minkowski's.
    \item $\chi(0) = 0$, so that boundary temperature corresponds to the one computed in the bulk's deep interior.
\end{enumerate}
The first condition can be imposed algorithmically by creating a feedback loop in the numerical integration of the equations of motion -- thus fixing one of the initial values, in our case the electric field on the horizon $A_t'(u_h)$ --; the second, instead, can be attained by defining an auxiliary field: $\eta(u) = \Omega - h(u)$, so that $\Omega$ becomes a quantity to be read on the UV boundary,making it trivial to impose $h(0)=0$; fixing $\eta(u_h)$, then, imposes condition 2 and avoids the need to specify $\Omega$ before integrating the equations of motion.

Additionally, we learn that the value of $\chi(u_h)$ is totally irrelevant, as a suitable rescaling of the time coordinate $t$ will do the trick of setting $\chi(0)=0$. In other terms: we are allowed to start with an arbitrary value of $\chi(u_h)$, then to rescale the time coordinate. This will shift the value of $\chi(u_h)$ to the one realizing $\chi(0)=0$.

At last, it is possible to notice that the two quantities $L_0$ and $\alpha$ appear, in the equations of motion, always in the combination $L_0/\alpha$, making them virtually indistinguishable. This does not come as totally unexpected: being $L_0$ a generic length scale, and $\alpha$ a parameter describing shape-preserving deformations, it makes sense that they are tied together. This allows us to define a single scale length:
\begin{equation}
    L_\alpha \equiv \frac{L_0}{\alpha},
\end{equation}
that substitutes the two distinct parameters $L_0$ and $\alpha$. After this few considerations, then, the map \eqref{eq:solution_map1} becomes:
\begin{equation}\label{eq:solution_map2}
    \left(\psi(u_h),\, \eta(u_h),\, u_h;\, L_\alpha \right) \longrightarrow \text{ Bulk fields},
\end{equation}
We can do better, and give a boundary-oriented physical meaning to this map.  By making use of these relations, the map \eqref{eq:solution_map2}, as seen from the boundary theory, is then:
\begin{equation}\label{eq:solution_map3}
    \left(T,\, \rho, \,\Omega; \,L_\alpha \right) \longrightarrow \text{ Bulk fields}.
\end{equation}
This means that once we fix the temperature, charge density and deformation state of the system, we are provided with a unique set of profiles for the bulk fields. In this, we should remember that the boundary theory still retains a property of scale invariance, so that the system is invariant under rescalings:
\begin{equation}
    T \rightarrow a T, \qquad \rho \rightarrow a^2 \rho, \qquad L_\alpha \rightarrow a^{-1} L_\alpha,
\end{equation}
as fixed by the dimensionality of these parameters.

All in all, a good choice to identify a state in the dual field theory is given by:
\begin{equation}
    \{T L_\alpha, \, \rho L_\alpha^2, \, \Omega\}
\end{equation}
which corresponds to the choice made in the main text.

\section{The effects of the charge density on the shear response}\label{app3}

For completeness, in this Appendix, we report some results which regard the effect of the background charge density $\rho$ on the mechanical response of the holographic system to shear deformations. In particular, we are interested in the behavior of the shear modulus and the shear viscosity in the solid phase in function of the charge density. Despite these same results have been used in the previous literature \cite{Ammon:2020xyv,Baggioli:2020edn}, we find useful to explicitly show them again here. Additionally, we also consider the full nonlinear response at finite charge density which, to the best of our knowledge, has not been considered explicitly in the literature.\medskip

\begin{figure}
\begin{subfigure}{.5\linewidth}
\centering
\includegraphics[trim=25 0 0 0,clip,width=0.9\linewidth]{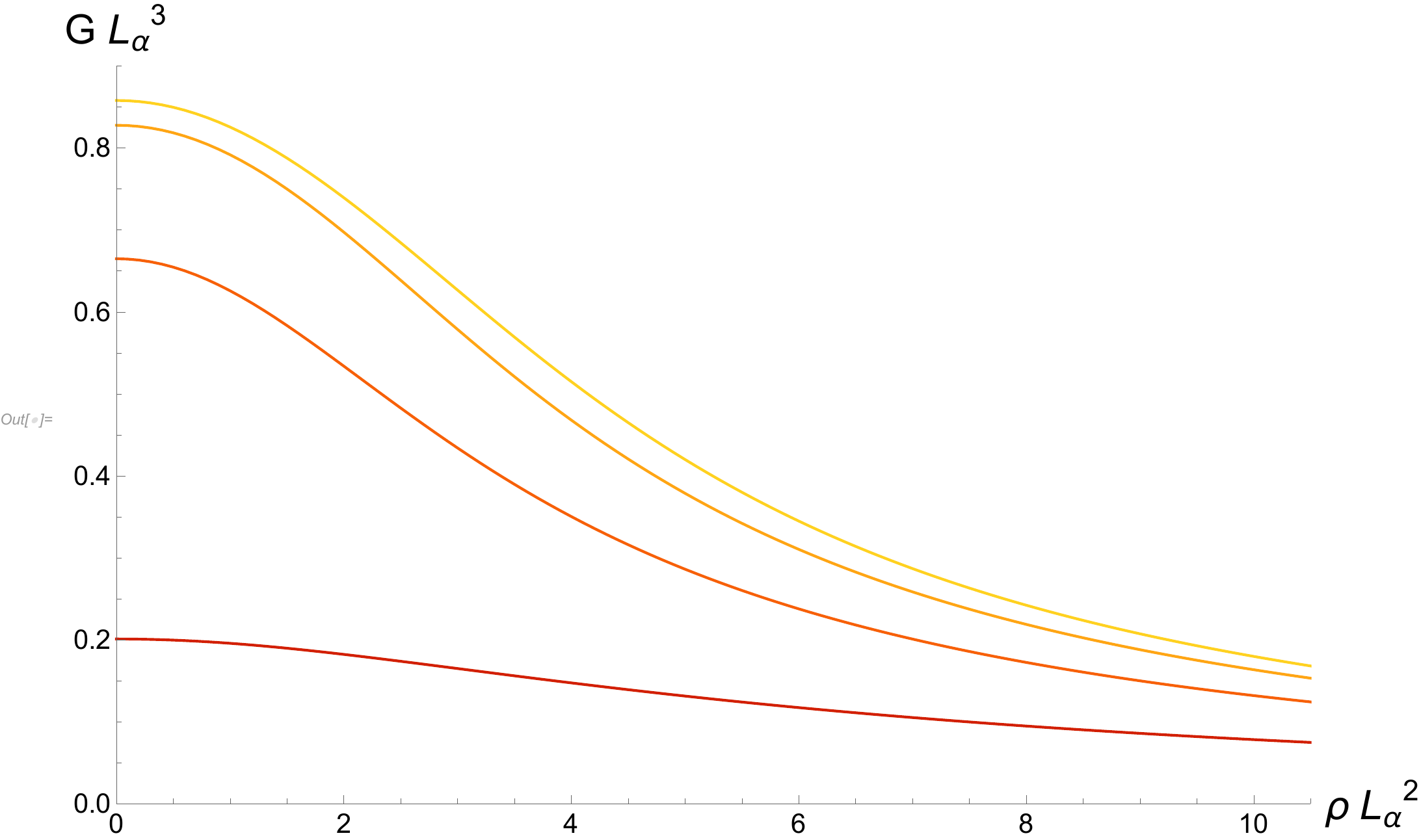}
\label{fig:sub1}
\end{subfigure}
\begin{subfigure}{.5\linewidth}
\centering
\includegraphics[trim=25 0 0 0,clip,width=0.9\linewidth]{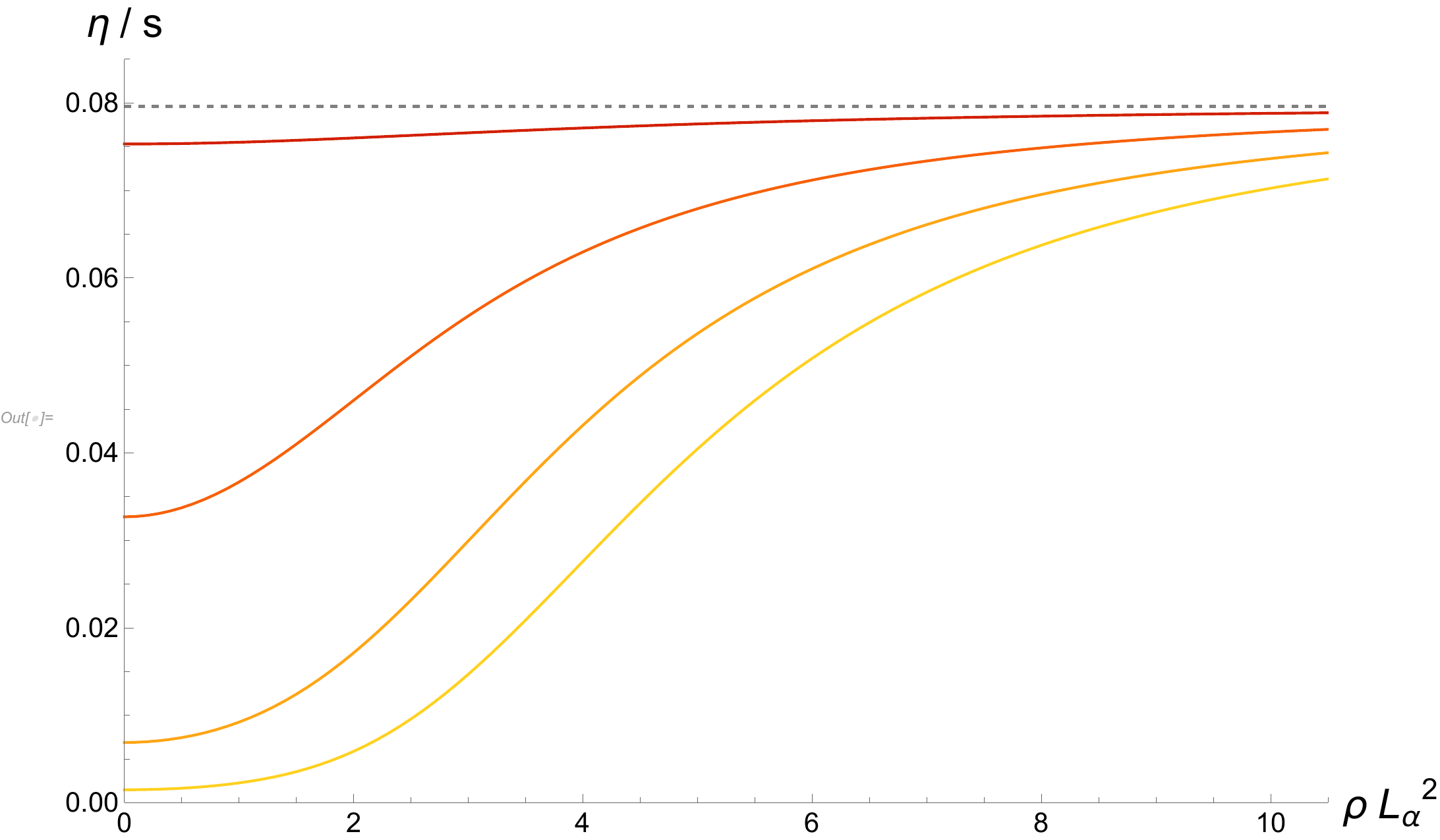}
\label{fig:sub2}
\end{subfigure}
\caption{Adimensional shear modulus (\textbf{Left}) and viscosity-over-entropy ratio (\textbf{Right}), for a set of adimensional temperatures: $T L_\alpha = 0.05,\, 0.1, \, 0.2,\, 0.4$ (from yellow to red).}
\label{fig:app_lin}
\end{figure}

In Fig.\ref{fig:app_lin}, we show the (adimensional) shear modulus and the viscosity-over-entropy ratio as a function of the charge density in the normal (solid) phase, at fixed temperatures. From there, we notice that the presence of finite charge density universally decreases the shear modulus and increases the value of the $\eta/s$ ratio towards its KSS value. The effects of the charge density are more pronounced for small temperatures and they become very mild at high temperature.\medskip

\begin{figure}
\begin{subfigure}{\linewidth}
\centering
\includegraphics[trim=25 0 0 0,clip,width=0.6\linewidth]{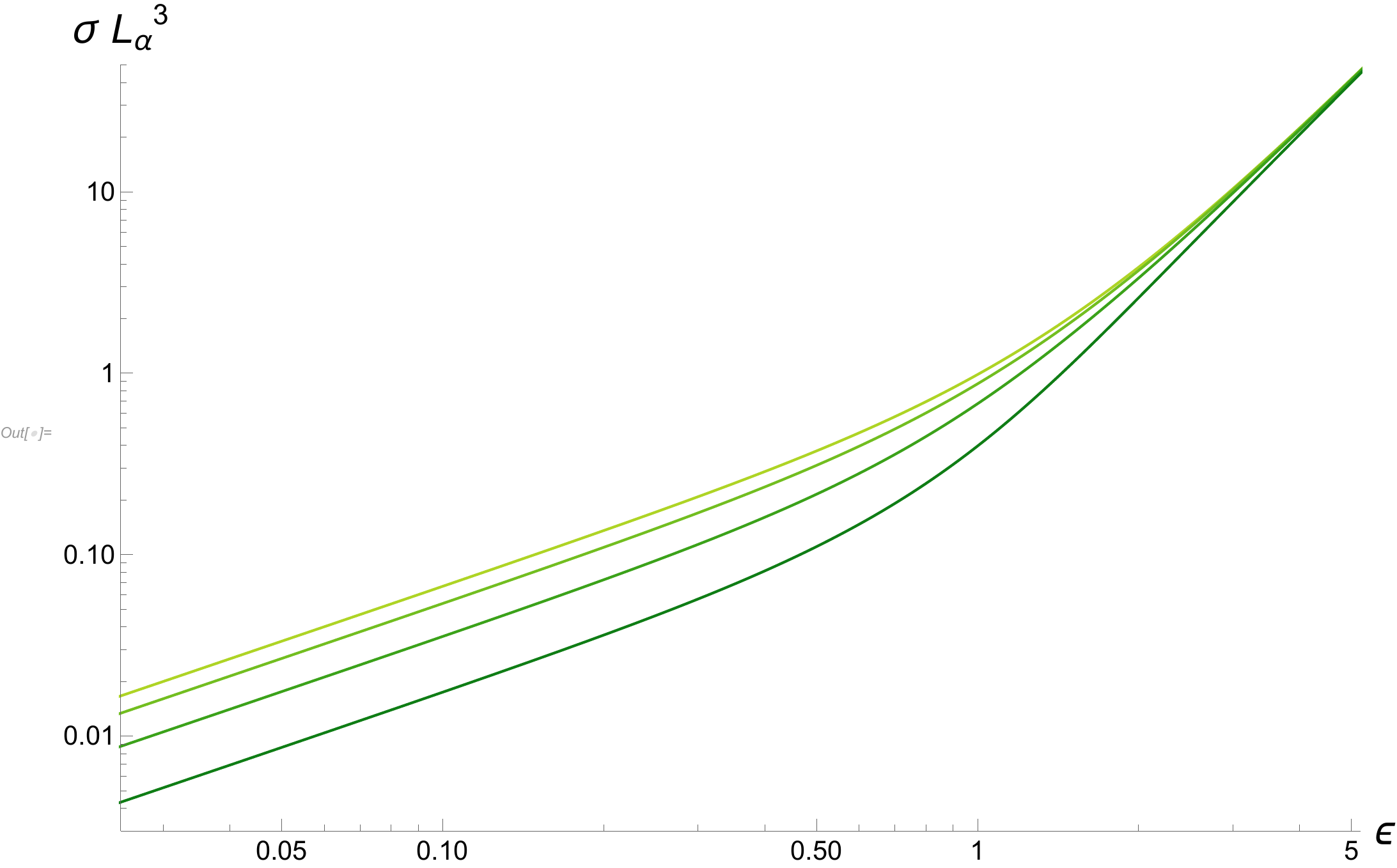}
\label{fig:sub3}
\end{subfigure}
\caption{Adimensional stress response as a function of the shear strain parameter $\epsilon$, at fixed adimensional temperature ($T L_\alpha=0.2$) for a set of increasing (adimensional) charge densities: $\rho L_\alpha^2 = 0,\, 2,\,4,\,8$ (from lighter to darker).}
\label{fig:app_nonlin}
\end{figure}

In Fig.\ref{fig:app_nonlin}, we focus on how the charge density affects the isothermal stress response of the system in the nonlinear regime. We find that the effects of the charge density are important only for small values of the strain and are consistent with the previous discussion about the shear modulus. In general, we see that the presence of charge density diminishes the elastic response of our system and that the effect is more important in the linear regime, i.e. at small shear strain.

\bibliographystyle{JHEP}
\bibliography{strain}

\providecommand{\href}[2]{#2}\begingroup\raggedright\begin{thebibliography}{100}

\bibitem{chaikin_lubensky_1995}
P.~M. Chaikin and T.~C. Lubensky, {\em Principles of Condensed Matter Physics}.
\newblock Cambridge University Press, 1995.

\bibitem{Leutwyler:1996er}
H.~Leutwyler, {\it {Phonons as Goldstone bosons}},  {\em Helv. Phys. Acta} {\bf
  70} (1997) 275--286, [\href{http://arxiv.org/abs/hep-ph/9609466}{{\tt
  hep-ph/9609466}}].

\bibitem{RevModPhys.71.S318}
A.~J. Leggett, {\it Superfluidity},  {\em Rev. Mod. Phys.} {\bf 71} (Mar, 1999)
  S318--S323.

\bibitem{Son:2002zn}
D.~T. Son, {\it {Low-energy quantum effective action for relativistic
  superfluids}},  \href{http://arxiv.org/abs/hep-ph/0204199}{{\tt
  hep-ph/0204199}}.

\bibitem{Nicolis:2011cs}
A.~Nicolis, {\it {Low-energy effective field theory for finite-temperature
  relativistic superfluids}},  \href{http://arxiv.org/abs/1108.2513}{{\tt
  arXiv:1108.2513}}.

\bibitem{doi:10.1063/1.3248499}
R.~J. Donnelly, {\it The two-fluid theory and second sound in liquid helium},
  {\em Physics Today} {\bf 62} (2009), no.~10 34--39,
  [\href{http://arxiv.org/abs/https://doi.org/10.1063/1.3248499}{{\tt
  https://doi.org/10.1063/1.3248499}}].

\bibitem{RevModPhys.84.759}
M.~Boninsegni and N.~V. Prokof'ev, {\it Colloquium: Supersolids: What and where
  are they?},  {\em Rev. Mod. Phys.} {\bf 84} (May, 2012) 759--776.

\bibitem{Balibar2010}
S.~Balibar, {\it The enigma of supersolidity},  {\em Nature} {\bf 464} (Mar,
  2010) 176--182.

\bibitem{PhysRev.106.161}
E.~P. Gross, {\it Unified theory of interacting bosons},  {\em Phys. Rev.} {\bf
  106} (Apr, 1957) 161--162.

\bibitem{GROSS195857}
E.~Gross, {\it Classical theory of boson wave fields},  {\em Annals of Physics}
  {\bf 4} (1958), no.~1 57--74.

\bibitem{RevModPhys.34.694}
C.~N. Yang, {\it Concept of off-diagonal long-range order and the quantum
  phases of liquid {He} and of superconductors},  {\em Rev. Mod. Phys.} {\bf
  34} (Oct, 1962) 694--704.

\bibitem{PhysRevA.2.256}
G.~V. Chester, {\it {Speculations on Bose-Einstein Condensation and Quantum
  Crystals}},  {\em Phys. Rev. A} {\bf 2} (Jul, 1970) 256--258.

\bibitem{PhysRevLett.25.1543}
A.~J. Leggett, {\it {Can a Solid Be "Superfluid"?}},  {\em Phys. Rev. Lett.}
  {\bf 25} (Nov, 1970) 1543--1546.

\bibitem{PhysRev.104.576}
O.~Penrose and L.~Onsager, {\it {Bose-Einstein Condensation and Liquid
  Helium}},  {\em Phys. Rev.} {\bf 104} (Nov, 1956) 576--584.

\bibitem{doi:10.1126/science.1196409}
H.~Choi, D.~Takahashi, K.~Kono, and E.~Kim, {\it {Evidence of Supersolidity in
  Rotating Solid Helium}},  {\em Science} {\bf 330} (2010), no.~6010
  1512--1515,
  [\href{http://arxiv.org/abs/https://www.science.org/doi/pdf/10.1126/science.1196409}{{\tt
  https://www.science.org/doi/pdf/10.1126/science.1196409}}].

\bibitem{kim2004probable}
E.~Kim and M.~Chan, {\it Probable observation of a supersolid helium phase},
  {\em Nature} {\bf 427} (2004), no.~6971 225--227.

\bibitem{kim2004observation}
E.~Kim and M.~H. Chan, {\it Observation of superflow in solid helium},  {\em
  Science} {\bf 305} (2004), no.~5692 1941--1944.

\bibitem{rittner2006observation}
A.~S.~C. Rittner and J.~D. Reppy, {\it Observation of classical rotational
  inertia and nonclassical supersolid signals in solid {H}e\textsubscript{4}
  below 250 m{K}},  {\em Phys. Rev. Lett.} {\bf 97} (Oct, 2006) 165301.

\bibitem{PhysRevLett.109.155301}
D.~Y. Kim and M.~H.~W. Chan, {\it {Absence of Supersolidity in Solid Helium in
  Porous Vycor Glass}},  {\em Phys. Rev. Lett.} {\bf 109} (Oct, 2012) 155301.

\bibitem{kim2014upper}
D.~Y. Kim and M.~H. Chan, {\it {Upper limit of supersolidity in solid helium}},
   {\em Physical Review B} {\bf 90} (2014), no.~6 064503.

\bibitem{doi:10.1126/science.aba4309}
L.~Tanzi, J.~G. Maloberti, G.~Biagioni, A.~Fioretti, C.~Gabbanini, and
  G.~Modugno, {\it {Evidence of superfluidity in a dipolar supersolid from
  nonclassical rotational inertia}},  {\em Science} {\bf 371} (2021), no.~6534
  1162--1165,
  [\href{http://arxiv.org/abs/https://www.science.org/doi/pdf/10.1126/science.aba4309}{{\tt
  https://www.science.org/doi/pdf/10.1126/science.aba4309}}].

\bibitem{Guo2021}
Y.~Guo, R.~M. Kroeze, B.~P. Marsh, S.~Gopalakrishnan, J.~Keeling, and B.~L.
  Lev, {\it An optical lattice with sound},  {\em Nature} {\bf 599} (Nov, 2021)
  211--215.

\bibitem{Norcia2021}
M.~A. Norcia, C.~Politi, L.~Klaus, E.~Poli, M.~Sohmen, M.~J. Mark, R.~N.
  Bisset, L.~Santos, and F.~Ferlaino, {\it Two-dimensional supersolidity in a
  dipolar quantum gas},  {\em Nature} {\bf 596} (Aug, 2021) 357--361.

\bibitem{PhysRevLett.104.075302}
J.~Day, O.~Syshchenko, and J.~Beamish, {\it {Nonlinear Elastic Response in
  Solid Helium: Critical Velocity or Strain?}},  {\em Phys. Rev. Lett.} {\bf
  104} (Feb, 2010) 075302.

\bibitem{doi:10.1063/1.2883895}
J.~Miller, {\it {Supersolid behavior in helium coincides with enhanced shear
  modulus}},  {\em Physics Today} {\bf 61} (2008), no.~2 14--15,
  [\href{http://arxiv.org/abs/https://doi.org/10.1063/1.2883895}{{\tt
  https://doi.org/10.1063/1.2883895}}].

\bibitem{Syshchenko_2009}
O.~Syshchenko, J.~Day, and J.~Beamish, {\it {Elastic properties of solid
  helium}},  {\em Journal of Physics: Condensed Matter} {\bf 21} (mar, 2009)
  164204.

\bibitem{Day2007}
J.~Day and J.~Beamish, {\it {Low-temperature shear modulus changes in solid
  He\textsubscript{4} and connection to supersolidity}},  {\em Nature} {\bf
  450} (Dec, 2007) 853--856.

\bibitem{prokof2007makes}
N.~Prokof'ev, {\it {What makes a crystal supersolid?}},  {\em Advances in
  Physics} {\bf 56} (2007), no.~2 381--402.

\bibitem{lin2009heat}
X.~Lin, A.~Clark, Z.~Cheng, and M.~Chan, {\it Heat capacity peak in solid
  {H}e\textsubscript{4}: Effects of disorder and {H}e\textsubscript{3}
  impurities},  {\em Physical review letters} {\bf 102} (2009), no.~12 125302.

\bibitem{anderson2008bose}
P.~W. Anderson, {\it {Bose Fluids Above T\textsubscript{c}: Incompressible
  Vortex Fluids and “Supersolidity”}},  {\em Physical review letters} {\bf
  100} (2008), no.~21 215301.

\bibitem{PhysRevB.53.5670}
H.~T.~C. Stoof, K.~Mullen, M.~Wallin, and S.~M. Girvin, {\it Hydrodynamics of
  spatially ordered superfluids},  {\em Phys. Rev. B} {\bf 53} (Mar, 1996)
  5670--5682.

\bibitem{Saslow2012}
W.~M. Saslow, {\it On the superfluid fraction and the hydrodynamics of
  supersolids},  {\em Journal of Low Temperature Physics} {\bf 169} (Nov, 2012)
  248--263.

\bibitem{hofmann2021hydrodynamics}
J.~Hofmann and W.~Zwerger, {\it Hydrodynamics of a superfluid smectic},  {\em
  Journal of Statistical Mechanics: Theory and Experiment} {\bf 2021} (2021),
  no.~3 033104.

\bibitem{PhysRevLett.97.125302}
J.~Ye, {\it Ginzburg-{L}andau theory of a supersolid},  {\em Phys. Rev. Lett.}
  {\bf 97} (Sep, 2006) 125302.

\bibitem{PhysRevLett.96.055301}
A.~T. Dorsey, P.~M. Goldbart, and J.~Toner, {\it Squeezing superfluid from a
  stone: Coupling superfluidity and elasticity in a supersolid},  {\em Phys.
  Rev. Lett.} {\bf 96} (Feb, 2006) 055301.

\bibitem{PhysRevLett.94.175301}
D.~T. Son, {\it {Effective Lagrangian and Topological Interactions in
  Supersolids}},  {\em Phys. Rev. Lett.} {\bf 94} (May, 2005) 175301.

\bibitem{RevModPhys.46.705}
C.~P. Enz, {\it Two-fluid hydrodynamic description of ordered systems},  {\em
  Rev. Mod. Phys.} {\bf 46} (Oct, 1974) 705--753.

\bibitem{Nicolis:2013lma}
A.~Nicolis, R.~Penco, and R.~A. Rosen, {\it {Relativistic Fluids, Superfluids,
  Solids and Supersolids from a Coset Construction}},  {\em Phys. Rev.} {\bf
  D89} (2014), no.~4 045002, [\href{http://arxiv.org/abs/1307.0517}{{\tt
  arXiv:1307.0517}}].

\bibitem{Delacretaz:2014jka}
L.~V. Delacr\'etaz, A.~Nicolis, R.~Penco, and R.~A. Rosen, {\it {Wess-Zumino
  Terms for Relativistic Fluids, Superfluids, Solids, and Supersolids}},  {\em
  Phys. Rev. Lett.} {\bf 114} (2015), no.~9 091601,
  [\href{http://arxiv.org/abs/1403.6509}{{\tt arXiv:1403.6509}}].

\bibitem{LVEFT3}
A.~Nicolis, R.~Penco, F.~Piazza, and R.~Rattazzi, {\it {Zoology of condensed
  matter: Framids, ordinary stuff, extra-ordinary stuff}},  {\em JHEP} {\bf 06}
  (2015) 155, [\href{http://arxiv.org/abs/1501.03845}{{\tt arXiv:1501.03845}}].

\bibitem{Krichevsky:2020ury}
R.~Krichevsky, {\em {Low-energy dynamics of condensed matter from the
  high-energy point of view: Studies in the effective field theory of matter}}.
\newblock PhD thesis, Columbia U. (main), 2020.

\bibitem{Celoria:2017bbh}
M.~Celoria, D.~Comelli, and L.~Pilo, {\it {Fluids, Superfluids and Supersolids:
  Dynamics and Cosmology of Self Gravitating Media}},  {\em JCAP} {\bf 09}
  (2017) 036, [\href{http://arxiv.org/abs/1704.00322}{{\tt arXiv:1704.00322}}].

\bibitem{zaanen2015holographic}
J.~Zaanen, Y.~Liu, Y.~Sun, and K.~Schalm, {\em Holographic Duality in Condensed
  Matter Physics}.
\newblock Cambridge University Press, 2015.

\bibitem{Natsuume:2014sfa}
M.~Natsuume, {\it {AdS/CFT Duality User Guide}},  {\em Lect. Notes Phys.} {\bf
  903} (2015) pp.1--294, [\href{http://arxiv.org/abs/1409.3575}{{\tt
  arXiv:1409.3575}}].

\bibitem{Hartnoll:2016apf}
S.~A. Hartnoll, A.~Lucas, and S.~Sachdev, {\it {Holographic quantum matter}},
  \href{http://arxiv.org/abs/1612.07324}{{\tt arXiv:1612.07324}}.

\bibitem{Baggioli:2019rrs}
M.~Baggioli, {\em {Applied Holography}: {A Practical Mini-Course}}.
\newblock SpringerBriefs in Physics. Springer, 2019.

\bibitem{Hartnoll:2008vx}
S.~A. Hartnoll, C.~P. Herzog, and G.~T. Horowitz, {\it {Building a Holographic
  Superconductor}},  {\em Phys. Rev. Lett.} {\bf 101} (2008) 031601,
  [\href{http://arxiv.org/abs/0803.3295}{{\tt arXiv:0803.3295}}].

\bibitem{Hartnoll_2008}
S.~A. Hartnoll, C.~P. Herzog, and G.~T. Horowitz, {\it Holographic
  superconductors},  {\em Journal of High Energy Physics} {\bf 2008} (dec,
  2008) 015--015.

\bibitem{Cai:2015cya}
R.-G. Cai, L.~Li, L.-F. Li, and R.-Q. Yang, {\it {Introduction to Holographic
  Superconductor Models}},  {\em Sci. China Phys. Mech. Astron.} {\bf 58}
  (2015), no.~6 060401, [\href{http://arxiv.org/abs/1502.00437}{{\tt
  arXiv:1502.00437}}].

\bibitem{Herzog:2011ec}
C.~P. Herzog, N.~Lisker, P.~Surowka, and A.~Yarom, {\it {Transport in
  holographic superfluids}},  {\em JHEP} {\bf 08} (2011) 052,
  [\href{http://arxiv.org/abs/1101.3330}{{\tt arXiv:1101.3330}}].

\bibitem{Schmitt:2014eka}
A.~Schmitt, {\it {Introduction to Superfluidity}: {Field-theoretical approach
  and applications}},  {\em Lect.Notes Phys.} {\bf 888} (2015) 1--155,
  [\href{http://arxiv.org/abs/1404.1284}{{\tt arXiv:1404.1284}}].

\bibitem{Arean:2021tks}
D.~Arean, M.~Baggioli, S.~Grieninger, and K.~Landsteiner, {\it {A holographic
  superfluid symphony}},  {\em JHEP} {\bf 11} (2021) 206,
  [\href{http://arxiv.org/abs/2107.08802}{{\tt arXiv:2107.08802}}].

\bibitem{Ammon:2021slb}
M.~Ammon, D.~Arean, M.~Baggioli, S.~Gray, and S.~Grieninger, {\it
  {Pseudo-spontaneous $U(1)$ Symmetry Breaking in Hydrodynamics and
  Holography}},  \href{http://arxiv.org/abs/2111.10305}{{\tt
  arXiv:2111.10305}}.

\bibitem{Donos:2021pkk}
A.~Donos, P.~Kailidis, and C.~Pantelidou, {\it {Dissipation in holographic
  superfluids}},  {\em JHEP} {\bf 09} (2021) 134,
  [\href{http://arxiv.org/abs/2107.03680}{{\tt arXiv:2107.03680}}].

\bibitem{baggioli2015electron}
M.~Baggioli and O.~Pujolas, {\it Electron-phonon interactions, metal-insulator
  transitions, and holographic massive gravity},  {\em Physical review letters}
  {\bf 114} (2015), no.~25 251602.

\bibitem{Baggioli:2021xuv}
M.~Baggioli, K.-Y. Kim, L.~Li, and W.-J. Li, {\it {Holographic Axion Model: a
  simple gravitational tool for quantum matter}},
  \href{http://arxiv.org/abs/2101.01892}{{\tt arXiv:2101.01892}}.

\bibitem{Donos:2013eha}
A.~Donos and J.~P. Gauntlett, {\it {Holographic Q-lattices}},  {\em JHEP} {\bf
  04} (2014) 040, [\href{http://arxiv.org/abs/1311.3292}{{\tt
  arXiv:1311.3292}}].

\bibitem{Grozdanov:2018ewh}
S.~Grozdanov and N.~Poovuttikul, {\it {Generalized global symmetries in states
  with dynamical defects: The case of the transverse sound in field theory and
  holography}},  {\em Phys. Rev. D} {\bf 97} (2018), no.~10 106005,
  [\href{http://arxiv.org/abs/1801.03199}{{\tt arXiv:1801.03199}}].

\bibitem{Nakamura:2009tf}
S.~Nakamura, H.~Ooguri, and C.-S. Park, {\it {Gravity Dual of Spatially
  Modulated Phase}},  {\em Phys. Rev. D} {\bf 81} (2010) 044018,
  [\href{http://arxiv.org/abs/0911.0679}{{\tt arXiv:0911.0679}}].

\bibitem{Amoretti:2017frz}
A.~Amoretti, D.~Arean, B.~Gouteraux, and D.~Musso, {\it {Effective holographic
  theory of charge density waves}},
  \href{http://arxiv.org/abs/1711.06610}{{\tt arXiv:1711.06610}}.

\bibitem{Alberte:2016xja}
L.~Alberte, M.~Baggioli, and O.~Pujolas, {\it {Viscosity bound violation in
  holographic solids and the viscoelastic response}},  {\em JHEP} {\bf 07}
  (2016) 074, [\href{http://arxiv.org/abs/1601.03384}{{\tt arXiv:1601.03384}}].

\bibitem{Alberte:2017oqx}
L.~Alberte, M.~Ammon, A.~Jim\'enez-Alba, M.~Baggioli, and O.~Pujol\`as, {\it
  {Holographic Phonons}},  {\em Phys. Rev. Lett.} {\bf 120} (2018), no.~17
  171602, [\href{http://arxiv.org/abs/1711.03100}{{\tt arXiv:1711.03100}}].

\bibitem{Baggioli:2019abx}
M.~Baggioli and S.~Grieninger, {\it {Zoology of solid \& fluid holography
  \textemdash{} Goldstone modes and phase relaxation}},  {\em JHEP} {\bf 10}
  (2019) 235, [\href{http://arxiv.org/abs/1905.09488}{{\tt arXiv:1905.09488}}].

\bibitem{Armas:2019sbe}
J.~Armas and A.~Jain, {\it {Viscoelastic hydrodynamics and holography}},  {\em
  JHEP} {\bf 01} (2020) 126, [\href{http://arxiv.org/abs/1908.01175}{{\tt
  arXiv:1908.01175}}].

\bibitem{Ammon:2020xyv}
M.~Ammon, M.~Baggioli, S.~Gray, S.~Grieninger, and A.~Jain, {\it {On the
  Hydrodynamic Description of Holographic Viscoelastic Models}},  {\em Phys.
  Lett. B} {\bf 808} (2020) 135691,
  [\href{http://arxiv.org/abs/2001.05737}{{\tt arXiv:2001.05737}}].

\bibitem{Armas:2020bmo}
J.~Armas and A.~Jain, {\it {Hydrodynamics for charge density waves and their
  holographic duals}},  \href{http://arxiv.org/abs/2001.07357}{{\tt
  arXiv:2001.07357}}.

\bibitem{Baggioli:2020edn}
M.~Baggioli, S.~Grieninger, and L.~Li, {\it {Magnetophonons \& type-B
  Goldstones from Hydrodynamics to Holography}},
  \href{http://arxiv.org/abs/2005.01725}{{\tt arXiv:2005.01725}}.

\bibitem{Kiritsis:2015oxa}
E.~Kiritsis and J.~Ren, {\it {On Holographic Insulators and Supersolids}},
  {\em JHEP} {\bf 09} (2015) 168, [\href{http://arxiv.org/abs/1503.03481}{{\tt
  arXiv:1503.03481}}].

\bibitem{Baggioli:2015zoa}
M.~Baggioli and M.~Goykhman, {\it {Phases of holographic superconductors with
  broken translational symmetry}},  {\em JHEP} {\bf 07} (2015) 035,
  [\href{http://arxiv.org/abs/1504.05561}{{\tt arXiv:1504.05561}}].

\bibitem{Baggioli:2015dwa}
M.~Baggioli and M.~Goykhman, {\it {Under The Dome: Doped holographic
  superconductors with broken translational symmetry}},  {\em JHEP} {\bf 01}
  (2016) 011, [\href{http://arxiv.org/abs/1510.06363}{{\tt arXiv:1510.06363}}].

\bibitem{Zeng:2014uoa}
H.~B. Zeng and J.-P. Wu, {\it {Holographic superconductors from the massive
  gravity}},  {\em Phys. Rev. D} {\bf 90} (2014), no.~4 046001,
  [\href{http://arxiv.org/abs/1404.5321}{{\tt arXiv:1404.5321}}].

\bibitem{Ling:2014laa}
Y.~Ling, P.~Liu, C.~Niu, J.-P. Wu, and Z.-Y. Xian, {\it {Holographic
  Superconductor on Q-lattice}},  {\em JHEP} {\bf 02} (2015) 059,
  [\href{http://arxiv.org/abs/1410.6761}{{\tt arXiv:1410.6761}}].

\bibitem{Andrade:2014xca}
T.~Andrade and S.~A. Gentle, {\it {Relaxed superconductors}},  {\em JHEP} {\bf
  06} (2015) 140, [\href{http://arxiv.org/abs/1412.6521}{{\tt
  arXiv:1412.6521}}].

\bibitem{Erdmenger:2015qqa}
J.~Erdmenger, B.~Herwerth, S.~Klug, R.~Meyer, and K.~Schalm, {\it {S-Wave
  Superconductivity in Anisotropic Holographic Insulators}},  {\em JHEP} {\bf
  05} (2015) 094, [\href{http://arxiv.org/abs/1501.07615}{{\tt
  arXiv:1501.07615}}].

\bibitem{Kim:2015dna}
K.-Y. Kim, K.~K. Kim, and M.~Park, {\it {A Simple Holographic Superconductor
  with Momentum Relaxation}},  {\em JHEP} {\bf 04} (2015) 152,
  [\href{http://arxiv.org/abs/1501.00446}{{\tt arXiv:1501.00446}}].

\bibitem{Jeong:2021wiu}
H.-S. Jeong and K.-Y. Kim, {\it {Homes' law in Holographic Superconductor with
  Gubser-Rocha model}},  \href{http://arxiv.org/abs/2112.01153}{{\tt
  arXiv:2112.01153}}.

\bibitem{Ling:2016lis}
Y.~Ling and X.~Zheng, {\it {Holographic superconductor with momentum relaxation
  and Weyl correction}},  {\em Nucl. Phys. B} {\bf 917} (2017) 1--18,
  [\href{http://arxiv.org/abs/1609.09717}{{\tt arXiv:1609.09717}}].

\bibitem{Kim:2016jjk}
K.-Y. Kim and C.~Niu, {\it {Homes' law in Holographic Superconductor with
  Q-lattices}},  {\em JHEP} {\bf 10} (2016) 144,
  [\href{http://arxiv.org/abs/1608.04653}{{\tt arXiv:1608.04653}}].

\bibitem{KimHomes}
K.-Y. Kim, K.~K. Kim, and M.~Park, {\it {Ward Identity and Homes' Law in a
  Holographic Superconductor with Momentum Relaxation}},
  \href{http://arxiv.org/abs/1604.06205}{{\tt arXiv:1604.06205}}.

\bibitem{Kiritsis:2015hoa}
E.~Kiritsis and L.~Li, {\it {Holographic Competition of Phases and
  Superconductivity}},  {\em JHEP} {\bf 01} (2016) 147,
  [\href{http://arxiv.org/abs/1510.00020}{{\tt arXiv:1510.00020}}].

\bibitem{baggioli2020black}
M.~Baggioli, O.~Pujol{\`a}s, et~al., {\it Black rubber and the non-linear
  elastic response of scale invariant solids},  {\em Journal of High Energy
  Physics} {\bf 2020} (2020), no.~9 1--36.

\bibitem{Ammon:2019wci}
M.~Ammon, M.~Baggioli, and A.~Jim\'enez-Alba, {\it {A Unified Description of
  Translational Symmetry Breaking in Holography}},  {\em JHEP} {\bf 09} (2019)
  124, [\href{http://arxiv.org/abs/1904.05785}{{\tt arXiv:1904.05785}}].

\bibitem{Alberte:2018doe}
L.~Alberte, M.~Baggioli, V.~C. Castillo, and O.~Pujolas, {\it {Elasticity
  bounds from Effective Field Theory}},  {\em Phys. Rev.} {\bf D100} (2019),
  no.~6 065015, [\href{http://arxiv.org/abs/1807.07474}{{\tt
  arXiv:1807.07474}}].

\bibitem{Pan:2021cux}
D.~Pan, T.~Ji, M.~Baggioli, L.~Li, and Y.~Jin, {\it {Non-linear elasticity,
  yielding and entropy in amorphous solids}},
  \href{http://arxiv.org/abs/2108.13124}{{\tt arXiv:2108.13124}}.

\bibitem{ZAMM:ZAMM19850650903}
R.~W. Ogden, {\em Non-Linear Elastic Deformations}.
\newblock WILEY-VCH Verlag, 1985.

\bibitem{Baggioli:2019elg}
M.~Baggioli, V.~C. Castillo, and O.~Pujolas, {\it {Scale invariant solids}},
  {\em Phys. Rev. D} {\bf 101} (2020), no.~8 086005,
  [\href{http://arxiv.org/abs/1910.05281}{{\tt arXiv:1910.05281}}].

\bibitem{Burikham:2016roo}
P.~Burikham and N.~Poovuttikul, {\it {Shear viscosity in holography and
  effective theory of transport without translational symmetry}},  {\em Phys.
  Rev. D} {\bf 94} (2016), no.~10 106001,
  [\href{http://arxiv.org/abs/1601.04624}{{\tt arXiv:1601.04624}}].

\bibitem{Hartnoll_bumpyBH}
S.~A. Hartnoll, D.~M. Ramirez, and J.~E. Santos, {\it Entropy production,
  viscosity bounds and bumpy black holes},  {\em Journal of High Energy
  Physics} {\bf 2016} (mar, 2016).

\bibitem{Ling:2016ien}
Y.~Ling, Z.-Y. Xian, and Z.~Zhou, {\it {Holographic Shear Viscosity in
  Hyperscaling Violating Theories without Translational Invariance}},  {\em
  JHEP} {\bf 11} (2016) 007, [\href{http://arxiv.org/abs/1605.03879}{{\tt
  arXiv:1605.03879}}].

\bibitem{Baggioli:2020ljz}
M.~Baggioli and W.-J. Li, {\it {Universal Bounds on Transport in Holographic
  Systems with Broken Translations}},  {\em SciPost Phys.} {\bf 9} (2020),
  no.~1 007, [\href{http://arxiv.org/abs/2005.06482}{{\tt arXiv:2005.06482}}].

\bibitem{Baggioli:2019jcm}
M.~Baggioli, M.~Vasin, V.~V. Brazhkin, and K.~Trachenko, {\it {Gapped momentum
  states}},  {\em Phys. Rept.} {\bf 865} (2020) 1--44,
  [\href{http://arxiv.org/abs/1904.01419}{{\tt arXiv:1904.01419}}].

\bibitem{Baggioli:2021ntj}
M.~Baggioli, M.~Landry, and A.~Zaccone, {\it {Deformations, relaxation and
  broken symmetries in liquids, solids and glasses: a unified topological field
  theory}},  \href{http://arxiv.org/abs/2101.05015}{{\tt arXiv:2101.05015}}.

\bibitem{Alberte:2015isw}
L.~Alberte, M.~Baggioli, A.~Khmelnitsky, and O.~Pujolas, {\it {Solid Holography
  and Massive Gravity}},  {\em JHEP} {\bf 02} (2016) 114,
  [\href{http://arxiv.org/abs/1510.09089}{{\tt arXiv:1510.09089}}].

\bibitem{Andrade:2019zey}
T.~Andrade, M.~Baggioli, and O.~Pujol\`as, {\it {Linear viscoelastic dynamics
  in holography}},  {\em Phys. Rev. D} {\bf 100} (2019), no.~10 106014,
  [\href{http://arxiv.org/abs/1903.02859}{{\tt arXiv:1903.02859}}].

\bibitem{Natsuume_shearvisc}
M.~Natsuume and M.~Ohta, {\it The shear viscosity of holographic superfluids},
  {\em Progress of Theoretical Physics} {\bf 124} (dec, 2010) 931--951.

\bibitem{PhysRevLett.108.175301}
S.~Saccani, S.~Moroni, and M.~Boninsegni, {\it Excitation spectrum of a
  supersolid},  {\em Phys. Rev. Lett.} {\bf 108} (Apr, 2012) 175301.

\bibitem{Cai:2017qdz}
R.-G. Cai, L.~Li, Y.-Q. Wang, and J.~Zaanen, {\it {Intertwined Order and
  Holography: The Case of Parity Breaking Pair Density Waves}},  {\em Phys.
  Rev. Lett.} {\bf 119} (2017), no.~18 181601,
  [\href{http://arxiv.org/abs/1706.01470}{{\tt arXiv:1706.01470}}].

\bibitem{Cremonini:2017usb}
S.~Cremonini, L.~Li, and J.~Ren, {\it {Intertwined Orders in Holography: Pair
  and Charge Density Waves}},  {\em JHEP} {\bf 08} (2017) 081,
  [\href{http://arxiv.org/abs/1705.05390}{{\tt arXiv:1705.05390}}].

\bibitem{Cremonini:2016rbd}
S.~Cremonini, L.~Li, and J.~Ren, {\it {Holographic Pair and Charge Density
  Waves}},  {\em Phys. Rev.} {\bf D95} (2017), no.~4 041901,
  [\href{http://arxiv.org/abs/1612.04385}{{\tt arXiv:1612.04385}}].

\bibitem{Amado:2013aea}
I.~Amado, D.~Are\'an, A.~Jim\'enez-Alba, K.~Landsteiner, L.~Melgar, and
  I.~Salazar~Landea, {\it {Holographic Superfluids and the Landau Criterion}},
  {\em JHEP} {\bf 02} (2014) 063, [\href{http://arxiv.org/abs/1307.8100}{{\tt
  arXiv:1307.8100}}].

\bibitem{Domenech:2010nf}
O.~Domenech, M.~Montull, A.~Pomarol, A.~Salvio, and P.~J. Silva, {\it {Emergent
  Gauge Fields in Holographic Superconductors}},  {\em JHEP} {\bf 08} (2010)
  033, [\href{http://arxiv.org/abs/1005.1776}{{\tt arXiv:1005.1776}}].

\bibitem{buehler1965effect}
E.~Buehler and H.~Levinstein, {\it {Effect of tensile stress on the transition
  Temperature and current-carrying capacity of Nb\textsubscript{3}Sn}},  {\em
  Journal of Applied Physics} {\bf 36} (1965), no.~12 3856--3860.

\bibitem{zhai2000effect}
H.~Zhai and W.~Chu, {\it {Effect of interfacial strain on critical temperature
  of YBa\textsubscript{2}Cu\textsubscript{3}O\textsubscript{7- $\delta$} thin
  films}},  {\em Applied Physics Letters} {\bf 76} (2000), no.~23 3469--3471.

\bibitem{medvedev2009electronic}
S.~Medvedev, T.~McQueen, I.~Troyan, T.~Palasyuk, M.~Eremets, R.~Cava,
  S.~Naghavi, F.~Casper, V.~Ksenofontov, G.~Wortmann, et~al., {\it Electronic
  and magnetic phase diagram of {$\beta$-Fe\textsubscript{1.01}Se} with
  superconductivity at 36.7 {K} under pressure},  {\em Nature materials} {\bf
  8} (2009), no.~8 630--633.

\bibitem{han2010superconductivity}
Y.~Han, W.~Li, L.~Cao, X.~Wang, B.~Xu, B.~Zhao, Y.~Guo, and J.~Yang, {\it
  Superconductivity in iron telluride thin films under tensile stress},  {\em
  Physical review letters} {\bf 104} (2010), no.~1 017003.

\bibitem{qing2012interface}
W.~Qing-Yan, L.~Zhi, Z.~Wen-Hao, Z.~Zuo-Cheng, Z.~Jin-Song, L.~Wei, D.~Hao,
  O.~Yun-Bo, D.~Peng, C.~Kai, et~al., {\it {Interface-induced high-temperature
  superconductivity in single unit-cell FeSe films on SrTiO\textsubscript{3}}},
   {\em Chinese Physics Letters} {\bf 29} (2012), no.~3 037402.

\bibitem{hicks2014strong}
C.~W. Hicks, D.~O. Brodsky, E.~A. Yelland, A.~S. Gibbs, J.~A. Bruin, M.~E.
  Barber, S.~D. Edkins, K.~Nishimura, S.~Yonezawa, Y.~Maeno, et~al., {\it
  {Strong increase of T\textsubscript{c} of
  Sr\textsubscript{2}RuO\textsubscript{4} under both tensile and compressive
  strain}},  {\em Science} {\bf 344} (2014), no.~6181 283--285.

\bibitem{steppke2017strong}
A.~Steppke, L.~Zhao, M.~E. Barber, T.~Scaffidi, F.~Jerzembeck, H.~Rosner, A.~S.
  Gibbs, Y.~Maeno, S.~H. Simon, A.~P. Mackenzie, et~al., {\it {Strong peak in
  T\textsubscript{c} of Sr\textsubscript{2}RuO\textsubscript{4} under uniaxial
  pressure}},  {\em Science} {\bf 355} (2017), no.~6321.

\bibitem{ahadi2019enhancing}
K.~Ahadi, L.~Galletti, Y.~Li, S.~Salmani-Rezaie, W.~Wu, and S.~Stemmer, {\it
  {Enhancing superconductivity in SrTiO\textsubscript{3} films with strain}},
  {\em Science advances} {\bf 5} (2019), no.~4 eaaw0120.

\bibitem{ruf2021strain}
J.~P. Ruf, H.~Paik, N.~J. Schreiber, H.~P. Nair, L.~Miao, J.~K. Kawasaki, J.~N.
  Nelson, B.~D. Faeth, Y.~Lee, B.~H. Goodge, et~al., {\it Strain-stabilized
  superconductivity},  {\em Nature Communications} {\bf 12} (2021), no.~1 1--8.

\bibitem{ghini2021strain}
M.~Ghini, M.~Bristow, J.~C.~A. Prentice, S.~Sutherland, S.~Sanna, A.~A.
  Haghighirad, and A.~I. Coldea, {\it Strain tuning of nematicity and
  superconductivity in single crystals of fese},  {\em Phys. Rev. B} {\bf 103}
  (May, 2021) 205139.

\bibitem{locquet1998doubling}
J.-P. Locquet, J.~Perret, J.~Fompeyrine, E.~M{\"a}chler, J.~W. Seo, and
  G.~Van~Tendeloo, {\it {Doubling the critical temperature of
  La\textsubscript{1.9}Sr\textsubscript{0.1}CuO\textsubscript{4} using
  epitaxial strain}},  {\em Nature} {\bf 394} (1998), no.~6692 453--456.

\bibitem{malinowski2020suppression}
P.~Malinowski, Q.~Jiang, J.~J. Sanchez, J.~Mutch, Z.~Liu, P.~Went, J.~Liu,
  P.~J. Ryan, J.-W. Kim, and J.-H. Chu, {\it Suppression of superconductivity
  by anisotropic strain near a nematic quantum critical point},  {\em Nature
  Physics} {\bf 16} (2020), no.~12 1189--1193.

\end{thebibliography}\endgroup
\end{document}